\documentclass[a4paper]{JHEP3}
\usepackage{epsfig}
\usepackage{ulem}
\renewcommand{\sout}[1]{\if #1 \fi}

\title{Neutrino oscillations in magnetically driven supernova explosions}
\author{Shio Kawagoe\\ 
Division of Theoretical Astronomy, National Astronomical Observatory of Japan, 2-21-1, Osawa, Mitaka, Tokyo, 181-8588, Japan\\ E-mail:
 \email{shio.k@nao.ac.jp}}
\author{Tomoya Takiwaki\\Center for Computational Astrophysics, National Astronomical Observatory of Japan, 2-21-1, Osawa, Mitaka, Tokyo, 181-8588, Japan
\\ E-mail:
\email{takiwaki.tomoya@nao.ac.jp}}
\author{Kei Kotake\\ 
Division of Theoretical Astronomy/Center for Computational Astrophysics, National Astronomical Observatory of Japan, 2-21-1, Osawa, Mitaka, Tokyo, 181-8588, Japan
\\ E-mail: \email{kkotake@th.nao.ac.jp}}

\abstract
{
We investigate neutrino oscillations from core-collapse supernovae that 
 produce magnetohydrodynamic (MHD) explosions.
By calculating numerically the flavor conversion of neutrinos 
in the highly non-spherical envelope, we study how the explosion anisotropy has impacts on the 
emergent neutrino spectra through the Mikheyev-Smirnov-Wolfenstein effect.
  In the case of the inverted mass hierarchy with 
 a relatively large $\theta_{13}$ ($\sin^2 2\theta_{13} \gtrsim 10^{-3}$),
we show that survival probabilities of 
$\bar{\nu}_e$ and $\nu_e$ seen from the rotational axis of the MHD supernovae 
(i.e., polar direction), can be significantly different from those along 
the equatorial direction.
 The event numbers of $\bar{\nu}_e$ 
 observed from the polar direction are predicted to show steepest decrease, reflecting 
 the passage of the magneto-driven shock to the so-called high-resonance regions. 
 Furthermore we point out that such a shock effect, 
depending on the original neutrino spectra, 
appears also for the low-resonance regions, which could lead to a noticeable 
decrease in the $\nu_e$ signals. 
 This reflects a unique nature of the magnetic explosion featuring a very early 
shock-arrival to the resonance regions, which is in sharp contrast to the 
neutrino-driven delayed supernova models.
 Our results suggest that the two 
features in the $\bar{\nu}_e$ and $\nu_e$ signals, if visible to the 
Super-Kamiokande for a Galactic supernova,
could mark an observational signature of the 
magnetically driven explosions, presumably linked to the formation of 
 magnetars and/or long-duration gamma-ray bursts.}
\preprint{...} 

\begin{document}
\section{Introduction}
Current estimates of Galactic core-collapse supernovae rates predict one 
core-collapse supernova event every $\sim 40 \pm 10$ year 
 (e.g., \cite{ando_new}). When a massive star 
($ \gtrsim 10 M_{\odot}$) \cite{heger03} undergoes a core-collapse supernova 
explosion in our Galactic center, copious numbers of neutrinos are produced, 
some of which 
may be detected on the earth. Such supernova neutrinos will carry valuable information 
from deep inside the core. In fact, the detection of neutrinos from SN1987A 
(albeit in the Large Magellanic Cloud)
 paved the way for {\it Neutrino Astronomy}, 
an alternative to conventional astronomy by electromagnetic waves 
\cite{hirata1987,bionta1987}. Even though neutrino events from SN1987A 
were just two dozens, they have been studied extensively and allowed us to have a 
confidence that the basic picture of core-collapse supernova is correct 
(\cite{Sato-and-Suzuki,1988-Suzuki}  see 
\cite{raffelt_review} for a recent review).
Over the last decades,
significant progress has been made in a ground-based large water 
Cherenkov detector as Super-Kamiokande (SK) \cite{Totsuka92} and 
also in a liquid scintillator detector as KamLAND \cite{Suzuki99}.
If a supernova occurs in our Galactic center ($\sim 10$ kpc), 
 about 10,000 $\bar{\nu}_e$ events are estimated to be detected by SK  
\cite{totani-sato-del} (see also \cite{kamland-beacom}).
 Those successful neutrino detections 
are important not only to study the supernova physics but also to unveil 
 the nature of neutrinos itself such as the neutrino oscillation parameters  
the mass hierarchy, 
and neutrino mass
 \cite{dighe, Fogli4, Lunardini03, Dighe03a, Takahashi3,Fogli2, Beacom:1999}(e.g., \cite{kotake_rev} for a recent review).


The neutrino flux we receive on the Earth 
is sensitive to the profile of the matter that the neutrinos encounter on their path.
The supernova neutrinos interact with electrons when the neutrinos
propagate through stellar matter via the well-known Mikheyev-Smirnov-Wolfenstein (MSW)
 effect \cite{Takahashi3, mikheev1985, mikheev1986, balantekin}. 
 Such effects have been extensively investigated so far from various points of view, 
 with a focus such as on the progenitor 
dependence of the early neutrino burst \cite{Takahashi5, kachelriess2005}
and on the Earth matter effects 
(e.g., \cite{Lunardini, dighe2004}).

The neutrino conversion efficiency via the MSW effect depends 
 sharply on the density/electron fraction gradients, thus 
 sensitive to the discontinuity produced by 
the passage of the supernova shock.
It is rather recently that such shock effects 
 have been focused on \cite{ Lunardini03,Schirato02, Lunardini08} 
(see \cite{duan_rev} for a recent review).
  The time dependence of the events showing a decrease of the average energy of 
$\nu_e$ in the case of normal mass hierarchy (or $\bar{\nu}_e$ in the
case of inverted mass hierarchy), monitors the evolution of the density profile 
like a tomography,  thus could provide a powerful test of the mixing 
 angle and the mass hierarchy (e.g., 
\cite{Fogli2, Tomas, Takahashi1}). 
Very recently the flavor conversions driven by neutrino self-interactions are 
attracting great attention, because they can induce dramatic observable effects such as 
by the spectral split or swap (e.g., \cite{raffelt,duan,dasgup1,dasgup2} and see references
 therein). They are
 predicted to emerge as a distinct feature in their energy spectra 
(see \cite{duan_rev, duan2006} for reviews of the rapidly 
growing research field and collective references therein). 
 The resonant spin-flavor conversion has been also studied both analytically 
 (e.g., \cite{lim1988, akhmedov1988, Akhmedov03} 
and references therein) and numerically 
(e.g., \cite{Ando03b, Totani1996}), 
which induces flavor conversions between neutrinos and antineutrinos 
in magnetized supernova envelopes.
 Those important ingredients related to the flavor conversions in the supernova 
environment have been studied often one by one in each study without putting all the 
effects together, possibly in order to highlight the new ingredient. In this sense,
 all the studies mentioned above should be regarded as complimentary towards the precise 
 predictions of supernova neutrinos. 
 
 Here it should be mentioned that 
most of those rich phenomenology of supernova neutrinos 
have been based on the spherically symmetric models of core-collapse 
supernovae \cite{totani-sato-del,Lunardini03,Takahashi5,Lunardini08}. 
On the other hand, there are accumulating observations
 indicating that core-collapse supernovae are globally aspherical commonly 
(e.g., \cite{Wang01}). Pushed by them, various 
mechanisms have been explored thus far by supernova modelers to understand the 
 central engine, such as the neutrino-heating mechanism aided by 
 convection and the standing accretion shock instability 
(e.g., \cite{marek,Iwakami07} and references therein), 
magnetohydrodynamic (MHD) mechanism relying on the extraction of the rotational 
energy of rapidly rotating protoneutron stars via magnetic fields 
(e.g., \cite{kotake_mhd,Burrows07,Takiwaki09}, see \cite{kotake_rev} 
for a recent review and collective references therein), and the 
 acoustic mechanism relying on the acoustic energy deposition via oscillating 
protoneutron stars \cite{Burrows06}.
Apart from simple parametrization to mimic anisotropy and stochasticity of the shock
 (e.g., \cite{friedland, choubey}),
 there have been a very few studies focusing how those anisotropies 
obtained by the recent supernova simulations could affect the neutrino oscillations 
 (\cite{Tomas,kneller}). 
This is mainly due to the lack of multidimensional supernova models, 
which are generally too 
 computationally expensive to continue the simulations till the shock waves 
propagate outward until they affect neutrino transformations.

Here we study the neutrino oscillations in the case of the MHD
 explosions of core-collapse supernovae. In this case, the highly collimated 
 shock pushed by the strong 
 magnetic pressure can blow up the massive stars along the rotational axis
 \cite{kotake_mhd,Burrows07,Takiwaki09}. 
 Such explosions are considered to precede the formations of magnetars 
(e.g., \cite{kotake_mhd, Takiwaki09, Bucciantini}), presumably linked to the so-called 
 collapsars (e.g., \cite{MacFadyen99,Nagataki07,harikae}), which attract 
 great attention recently as a central engine of long-duration gamma-ray bursts (GRBs) 
(e.g., \cite{piran}). 
By performing special relativistic simulations  \cite{Takiwaki09}, it is recently 
 made possible to follow the MHD explosions in a consistent manner, starting 
from the onset of gravitational collapse, through
 core-bounce, the magnetic shock-revival, till the shock-propagation to the stellar 
 surface. Based on such models, we calculate numerically 
the flavor conversion in the highly non-spherical envelope 
through the MSW effect. For simplicity, we 
 neglect the effects of neutrino self-interactions and the resonant spin-flavor 
 conversion in this study.
 Even though far from comprehensive in this respect, 
we present the first discussion how the magneto-driven explosion 
anisotropy has impacts on the emergent neutrino spectra 
and the resulting event number observed by the 
SK for a future Galactic supernova.

The paper opens with generalities on magneto-driven explosion models and details on 
 our numerical model (section 2). In section 3, main results are shown, in
 which we pay attention 
 to the shock-effects on the so-called high-resonance in section 3.1-3.3 
 and on the low-resonance
 in section 3.4, respectively. In section 4, we give a discussion focusing on the 
 dependence of $\theta_{13}$ and also of the original 
neutrino spectra on the obtained results. Summary is given in section 5 with implications of our 
findings. 


\section{Numerical method}
\subsection{MHD driven supernova explosion model}

\FIGURE[t]{\epsfig{file=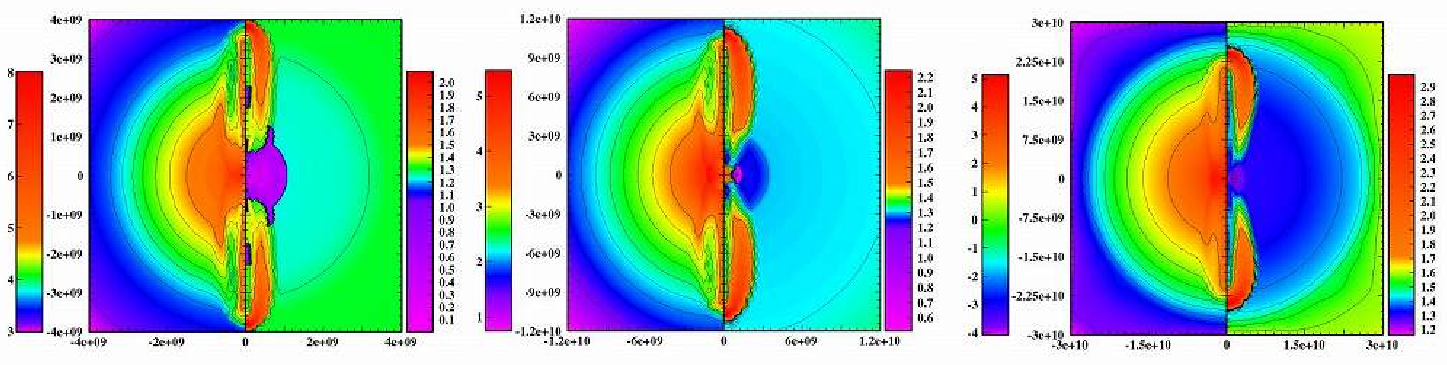}
\caption{ Snapshots showing MHD explosions of 
 core-collapse supernovae at 0.8, 2.0 and 4.3 sec after core bounce from
  left to right. In each figure, contour of the logarithmic density
  [$\mathrm{g/cm^3}$] (left) and entropy per baryon [$k_{\mathrm{B}}$]
  (right) are shown. The unit of the horizontal and the vertical axis is
  in [cm]. Note that the  difference of the length scale for each panel
  and that the outermost radius of the  progenitor is 
  $\sim 3 \times 10^{10}$ [cm]. }\label{fig:densentro}}


{We take time-dependent density profiles from our 2.5 dimensional MHD
  simulations of core-collapse supernovae \cite{Takiwaki09}, 
 in which a 25 M$_{\odot }$ presupernova model
by Heger et al. \cite{Heger} was adopted.
This progenitor lacks the hydrogen and helium layer during stellar 
 evolution, which is reconciled with observations that the progenitors 
 associated with long-duration gamma-ray bursts are type Ib/c supernovae 
(e.g., \cite{piran, Galama98}).

For our MHD model, the strong precollapse magnetic fields of 
$\sim 10^{12} \mathrm{G}$ are imposed with the high angular velocity of $\sim$ 72 ${\rm rad}/{\rm s}$ with a quadratic cutoff at 100km in radius, in which the rotational energy is equal
 to $\sim$1\% of the gravitational energy of the iron core (model B12TW1.0 in 
 \cite{Takiwaki09}).
 Such a combination of rapid rotation and 
 strong magnetic field, although strong differential rotation is assumed here, 
 have been often employed in the context of MHD supernova models
 \cite{kotake_mhd,Takiwaki09,Obergaulinger}.
 In those models, the explosion proceeds by the strong magnetic pressure 
amplified by the differential rotation near the surface of the protoneutron star.

One prominent feature of the MHD models is a high degree of the explosion asphericity. 
Figure \ref{fig:densentro} shows
 several snapshots featuring typical hydrodynamics of the model, 
from near core-bounce (0.8 s, left), during the 
 shock-propagation (2.0 s, middle), till near the shock break-out from the star (4.3 s, 
right), in which time is measured from the epoch of core bounce.
It can be seen that the strong shock propagates outwards with time along 
 the rotational axis.
On the other hand, the density profile hardly changes in the equatorial
direction, which is a generic feature of the MHD explosions observed in  
 recent numerical simulations \cite{kotake_mhd,Takiwaki09,Obergaulinger}.

As well known, the flavor conversion through the pure-matter MSW effect occurs 
in the resonance layer, where the density is 
\begin{eqnarray}
\rho _{\mathrm{res}} \sim  1.4 \times 10^3 \mathrm{g/cm}^3
 \left( \frac{\Delta m^2}{10^{-3} \mathrm{eV}^2}\right)
 \left( \frac{10 \mathrm{MeV}}{E_{\nu}} \right)
 \left( \frac{0.5}{Y_e} \right)
 \cos 2 \theta \label{eq:rho-res}
\end{eqnarray}
where $\Delta m^2$ is the mass squared difference, 
$E_{\nu}$ is the neutrino energy, ${Y_e}$ is the number of electrons per baryon,
and $\theta$ is the mixing angle.
 Since the inner supernova core is too dense to allow MSW resonance conversion, 
we focus on two resonance points in the outer supernova envelope.
One that occurs at higher density is called the H-resonance, and the other, 
 which occurs at lower density, is called the L-resonance. 
$\Delta m^2 $ and $\theta $ correspond to $\Delta m^2_{13}$ and
$\theta_{13}$  at the H-resonance and to $\Delta m^2_{12}$ and
$\theta_{12}$ at the L-resonance.

Figures \ref{fig:radH} and \ref{fig:radL} are evolutions of the density  
profiles in the polar direction every 0.4 s
as a function of radius.
Above and below horizontal lines show approximately density
of the resonance for different neutrino energies which are 5 and 60
MeV, respectively. 
Blue lines (left) 
show the range of the density of the H-resonance, and sky-blue lines (right) show 
that of the L-resonance. 
Along the polar axis, the shock wave reaches to the H-resonance,
$\sim O(10^3$)g/cm$^3$ at $\sim 0.5 $ s, and the L-resonance, $\sim O(1)$g/cm$^3$
 at $\sim 1.2$ s. It should be noted that 
those timescales are very early in comparison with the ones predicted in the 
neutrino-driven explosion models, typically $\sim $ 5 s and $\sim$ 
15 s for the H- and L- resonances,
 respectively (e.g., \cite{Fogli2, Tomas}). 
This arises from the fact that the MHD explosion 
is triggered promptly after core bounce without the shock-stall,
 which is in sharp contrast to the 
neutrino-driven {\it delayed} explosion models (\cite{Fogli2, Tomas}). 
The progenitor of the MHD 
models, possibly linked to long-duration gamma-ray bursts, is more compact 
due to the mass loss of hydrogen and helium envelopes 
during stellar evolution \cite{Woosley06},
 which is also the reason for the early shock-arrival to the resonance regions.

 Figure \ref{fig:radE} shows the density profiles in the equatorial direction 
every 0.4 s. The profiles are almost unchanged, because the MHD shock 
does not reach to the outer layer for the equatorial direction.  
It is noted in Figures 2 and 3 that to maximize the shock effect,
a sharpness of the shock along the polar direction 
is modified from the one in the MHD simulations, which 
is made to be inevitably blunt due to the employed numerical scheme using 
 an artificial viscosity to capture shocks \cite{Takiwaki09}. 
Green line in Figure \ref{fig:radH} is an original density profile for 2.0 s,
 whose shock front is replaced with the sharp vertical surface passing through the midpoint of 
the blunt shock.
 Although our simulations are one of the state-of-the-art MHD simulations at present, 
which can encompass the long-term hydrodynamics starting from gravitational collapse
 till the shock-breakout \cite{Takiwaki09}, this is really a limitation of the 
current MHD simulations.

Snapshots labeled as (a), (b), (c), (d), (e) and (f) in Figures \ref{fig:radH} and \ref{fig:radL} correspond to 0.4, 0.8 and 2.8, 1.2, 2.0 and 3.2 s after bounce, respectively,
 which is chosen for later convenience to compute the time evolutions of the 
flavor conversions.

\DOUBLEFIGURE[t]
  {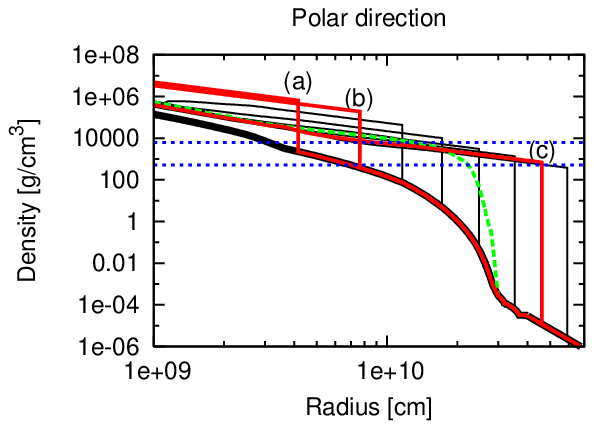}{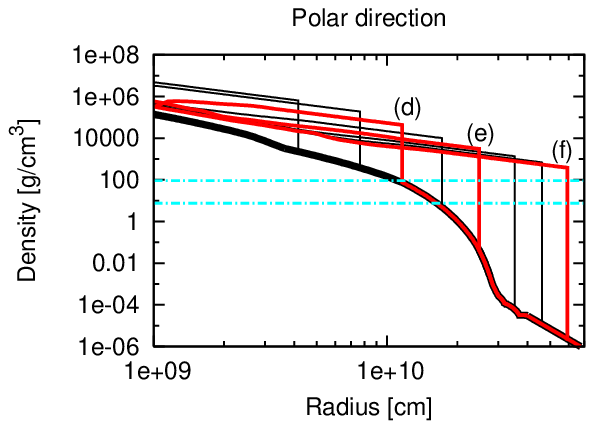}
  {\label{fig:radH}
  The density profile in the polar direction every 0.4 s as a
  function of radius. (a), (b) and (c) correspond to 0.4, 0.8 and 2.8
  s, respectively.   
  The horizontal blue lines show approximately density
  of the H-resonance for different neutrino energies which are 5 (above) and 60
  MeV (below), respectively. The green line is an original density
  profile for 2.0 s.
}
{\label{fig:radL}
Same as Figure 2, but for 1.2 (d), 2.0(e), and 3.2 (f)
  s, respectively.}

\DOUBLEFIGURE[t]
  {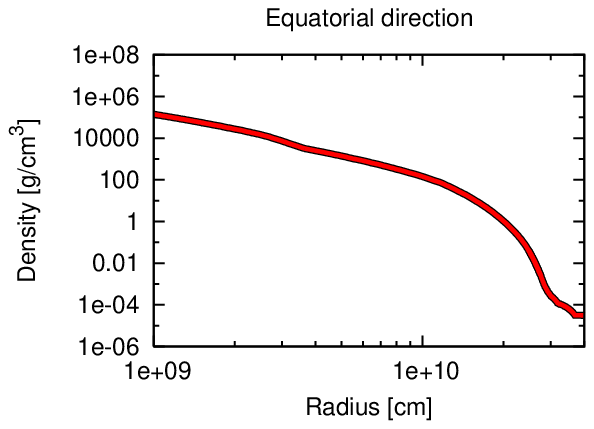}{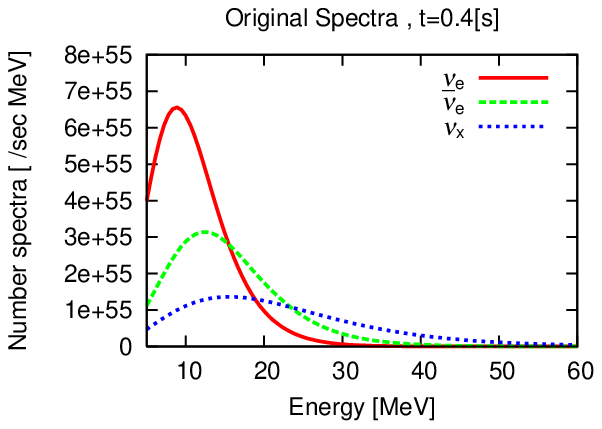}
{\label{fig:radE}
The density profile in equatorial direction every 0.4 s as a
  function of radius. 
The highly collimated shock does not propagate in the equatorial direction.
} 
{\label{fig:originalspect}The original spectra at 0.4 s
 after core bounce taken from the Lawrence Livermore simulation
 (e.g., \cite{totani-sato-del}). 
The solid line (red line), the dashed line (green) and
 the dotted line (blue) are $\nu_e$,
 $\bar{\nu}_e$ and $\nu_x$ spectra, respectively.}

\subsection{Neutrino oscillation calculations }\label{sn-neu-model}
 Along the time-dependent density profiles as shown in 
Figures \ref{fig:radH}-\ref{fig:radE}, we solve numerically the time evolution 
equation for the neutrino wave functions as follows,
\begin{eqnarray}
\imath \frac{d}{dt}
 \left( 
   \begin{array}{c}
     \nu_e \\ \nu_{\mu} \\ \nu_{\tau} 
   \end{array}
 \right)
=
U
 \left(
   \begin{array}{ccc}
     0  & 0           & 0 \\
     0  & \Delta E_{12} & 0 \\
     0  & 0           & \Delta E_{13} 
   \end{array}
 \right)
U^{-1}
 \left( 
   \begin{array}{c}
     \nu_e \\ \nu_{\mu} \\ \nu_{\tau} 
   \end{array}
 \right) 
+ 
 \left(
   \begin{array}{ccc}
     \sqrt{2} G_{\rm{F}}n_e & 0 & 0\\
     0  &  0  &  0  \\
     0  &  0  &  0 
   \end{array}
 \right)
 \left( 
   \begin{array}{c}
     \nu_e \\ \nu_{\mu} \\ \nu_{\tau} 
   \end{array}
 \right), 
\end{eqnarray}
where $\Delta E_{ij}=\Delta m^2_{ij}/2E_{\nu}$, and
$\Delta m^2_{ij}$ are  mass squared differences, $E_{\nu}$ is a
neutrino energy, $G_{\rm F}$ is the Fermi constant, 
and $n_e$ is an electron number density.
In the case of anti-neutrino, the sign of
 $\sqrt{2}\mathrm{G}_{F}n_e$ changes. 
$U$ is the  Cabibbo-Kobayashi-Masukawa (CKM) 
matrix,
\begin{eqnarray}
U=
\left(
 \begin{array}{ccc}
           c_{12}c_{13} & s_{12}c_{13} & s_{13}e^{-i\delta }\\
         -s_{12}c_{23}-c_{12}s_{23}s_{13}e^{i\delta } & 
         c_{12}c_{23}-s_{12}s_{23}s_{13}e^{i \delta } &
         s_{23}c_{13} \\
         s_{12}s_{23}-c_{12}c_{23}s_{13}e^{i\delta }  &
         -c_{12}s_{23}-s_{12}c_{23}s_{13}e^{i \delta } &
         c_{23}c_{13}
 \end{array}
\right),
\end{eqnarray}
where $s_{ij}=\sin {\theta}_{ij}$ and $c_{ij}=\cos {\theta}_{ij}$
($i\neq j =1,2,3$) are the mixing angle.
We set the CP violating phase, $\delta $, 
equal to zero in the CKM matrix for simplicity. 
 Solving numerically the above equation,
we obtain neutrino survival probabilities $P$ ($\bar{P}$)
that neutrinos emitted as $\nu_e$ ($\bar{\nu}_e$) at the neutrino
sphere remain $\nu_e$ ($\bar{\nu}_e$) at the surface of the
star \cite{neutrino}.
The survival probabilities are determined by an adiabaticity parameter
$\gamma$, 
\begin{eqnarray}
\gamma &=& 
\frac{\Delta m^2 \sin^2 2 \theta }{2 E_{\nu} \cos 2 \theta \left(1/n_e\right)(dn_e/dr)}.
\label{eq:gamma}
\end{eqnarray}
$\Delta m^2$ and $\theta$ correspond to $\Delta m^2_{13}$ and
$\theta_{13}$ at 
H-resonance, and to $\Delta m^2_{12}$ and $\theta_{12}$ at L-resonance,
respectively. 
When $\gamma \gg 1$, the resonance is called ``adiabatic'' and the conversion
between mass eigenstates does not occur.
On the other hand, when $\gamma \ll 1$,  the resonance is called ``non-adiabatic''
and the mass eigenstates are completely exchanged.
If the mass hierarchy is normal, the H- and L-resonances
occur only in the neutrino sector.
On the other hand, if the hierarchy is inverted, the H-resonance occurs in
 the anti-neutrino sector, and the L-resonance occurs in the neutrino sector.
The neutrino oscillation parameters are taken as 
$\sin ^2 2 \theta _{12}$=0.84,  $\sin ^2 2 \theta _{23}$=1.00,
$\Delta m^2_{12}$=8.1$\times 10^{-5}$eV$^2$ and $|\Delta
m^2_{13}|$=2.2$\times 10^{-3}$eV$^2$ (e.g., the summary in \cite{Parameter}
 and references therein).
For the unknown properties, we assume inverted mass hierarchy 
and $\sin^2 2 \theta _{13}$=1.0$\times$10$^{-3}$ as a fiducial value in 
our computations. This is because observable features in $\bar{\nu}_e$, 
most accessible channel to the SK, would imply that the neutrino mass hierarchy is
inverted and that $\theta_{13}$ is large.
The dependence of this parameter on the results will be discussed 
in Section \ref{dis-theta-13}.

The neutrino energy spectra at the surface of the star,
$\phi _{\nu}^{\mathrm{SN}}(E_{\nu})$, 
are calculated by multiplying the survival probabilities
by the original supernova neutrino spectra, $\phi ^{0}_{\nu}(E_{\nu})$ 
\cite{dighe},
\begin{eqnarray}
\left(
 \begin{array}{c}
  \phi ^{\mathrm{SN}}_{\nu_e}(E_{\nu}) \\
  \phi ^{\mathrm{SN}}_{\bar{\nu_e}}(E_{\nu}) \\
  \phi ^{\mathrm{SN}}_{\nu_x}(E_{\nu})
 \end{array}
\right)
=
\left(
 \begin{array}{ccc}
   P(E_{\nu}) & 0 & 1- P(E_{\nu}) \\
  0 & \bar{P}(E_{\nu})  & 1-\bar{P}(E_{\nu}) \\
  1-P(E_{\nu}) & 1-\bar{P}(E_{\nu}) &
                  2+ P(E_{\nu})+\bar{P}(E_{\nu})
\end{array}
\right)
\left(
 \begin{array}{c}
  \phi ^{0}_{\nu_e}(E_{\nu}) \\
  \phi ^{0}_{\bar{\nu_e}}(E_{\nu}) \\
  \phi ^{0}_{\nu_x}(E_{\nu})
 \end{array}
\right), \label{eq:spctra}
\end{eqnarray}
where
$\phi_{\nu_x} \equiv
1/4(\phi_{\nu_{\mu}}+\phi_{\nu_{\tau}}+\phi_{\bar{\nu}_{\mu}}
+\phi_{\bar{\nu}_{\tau}})$.
 To continue the simulations till the MHD shocks 
propagate outward until they affect neutrino transformations,
 the protoneutron stars are excised when the 
 shocks comes out of the iron core. By doing so, the severe 
 Courant-Friedrich-Levy condition in the stellar center can be relaxed,
 which is often taken in the long-term supernova simulation 
(e.g., \cite{Iwakami07,kneller} and references therein). 
A major drawback by the excision is that 
the information of the neutrino spectra at the neutrino sphere is lost.
 Assuming that the qualitative behavior in the neutrino luminosity, such as the 
 peaking near neutronization and the subsequent decay with time, is similar between the 
 neutrino-heating and the MHD mechanism,
 we employ the results of a full-scale 
 numerical simulation by the Lawrence Livermore group 
(e.g., as in \cite{totani-sato-del,Takahashi1}),
 which predict the neutrino temperatures of $\nu_{e}$, $\bar{\nu_{e}}$ and $\nu_{x}$
 to be 2.8, 4.0, and 7.0 MeV, respectively. 
The neutrino spectra at the neutrino 
sphere are approximated by Fermi-Dirac distributions and the neutrino luminosity is 
 taken to decay exponentially with a timescale of $t_{\rm L} = 3$ s 
 (e.g., \cite{yoshida2}). The total neutrino energy is assumed to be fixed at 
$2.9\times10^{53}\mathrm{erg}$
that corresponds to the model of the Lawrence Livermore group.
Figure \ref{fig:originalspect} shows the original spectra at 0.4 s after
core bounce. 
 The variation of the original neutrino spectra and its effects on our results will be
 discussed in Section \ref{dif-spe}.

We calculate expected event numbers of the supernova neutrinos at a water  
Cherenkov detector.
The positron (or electron) energy spectrum is evaluated as follows,
\begin{eqnarray}
\frac{d^2 N}{dE_e dt}=N_{\rm tar}\cdot \eta{(E_e)} \cdot
  \frac{1}{4 \pi d^2} \cdot \frac{d^2 N_{\nu}}{dE_{\nu }dt}\cdot
   \sigma(E_{\nu})\cdot \frac{dE_{\nu}}{dE_{e}}
\label{eq:ivent},
\end{eqnarray}
where $N$ is a detection number of neutrinos, $N_{\rm tar}$ is a target number, 
$\eta{(E_e)}$ is a efficiency of the detector,
$E_e$ is an energy of electron or positron, 
 $d$ is a distance from the supernova, 
$\frac{d^2N_{\nu}}{dE_{\nu}dt}$ is a neutrino
spectrum from the star calculated by the procedure mentioned above, and
$\sigma $ is the corresponding cross section given in \cite{neutrino}.
We consider the SK detector which 
is filled with 32000 ton pure water. 
The finite energy resolution of the detector is neglected here.
The event number is obtained by integrating over the angular
distribution of the events.
For simplification,
we take the efficiency of the detector as follows:
$\eta {(E_e)}=0$ at $E_e < 7$ [MeV] and
$\eta {(E_e)}=1$ at $E_e \geq 7 $ [MeV] \cite{threshold}.
We assume that the supernova occurs in our Galactic center ($d =$ 10 [kpc]). 
Albeit important \cite{Lunardini,dighe2004}, we do not consider the Earth matter 
 effect to highlight first the MSW effect inside the star.

\section{Result}
\subsection{Survival probability of $\bar{\nu}_e$}\label{Surv-prob-H}

\FIGURE[t]{
\epsfig{file=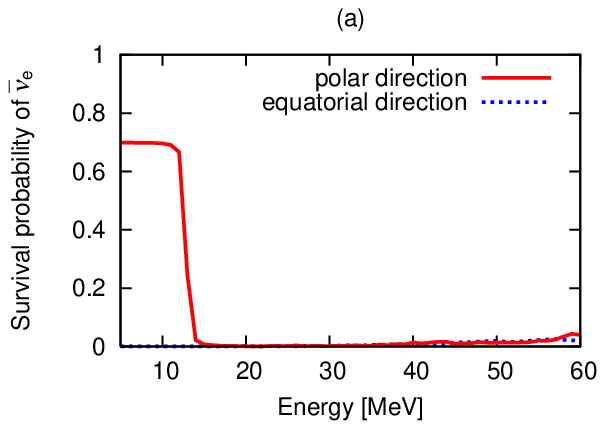,width=0.55\textwidth}
\epsfig{file=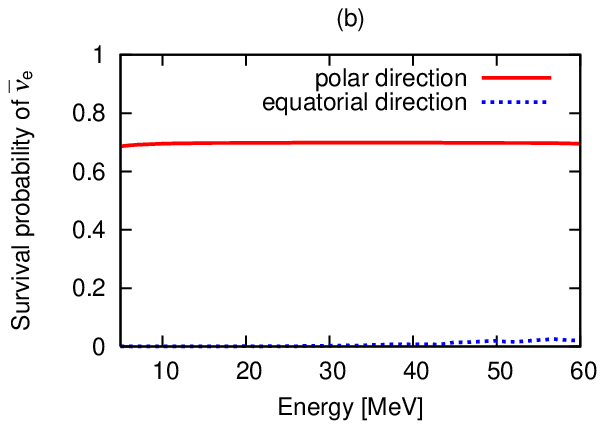,width=0.55\textwidth}
\epsfig{file=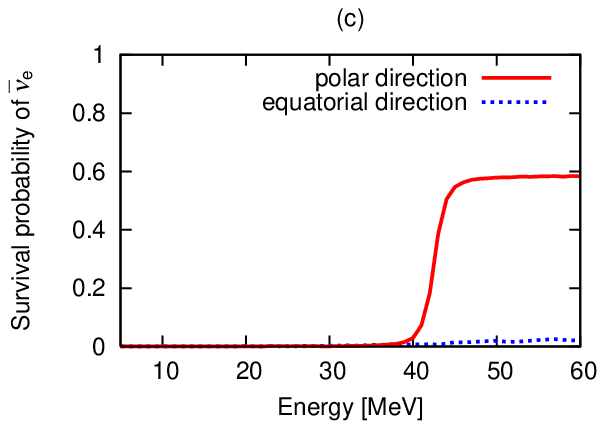,width=0.55\textwidth}
 \caption{\label{fig:surveb}The survival probability of $\bar{\nu}_e$ in
  the case of (a), (b) and (c). 
  The solid lines (red lines) and the dotted lines (blue lines) are for the polar
 and equatorial direction, respectively.  
Inverted mass hierarchy and $\sin ^2 2 \theta_{13}=10^{-3}$ are assumed.
  The survival probability of the polar direction changes from very
  early stage due to
  the faster shock-arrival to the resonance layers than the neutrino-driven supernova 
 models (See text for more details). }
}


In this section, we focus on
the influence of the shock wave in the H-resonance that appears in
the $\bar{\nu}_e$ signatures in the inverted mass hierarchy. 
First, we calculate survival probability of $\bar{\nu}_e$,
$\bar{P}(E_{\nu})$,  
in the manner described in Section \ref{sn-neu-model}.

Figure \ref{fig:surveb} is the survival probability of $\bar{\nu}_e$
as a function of neutrino energy in the case of (a), (b) and (c).
The solid lines (red lines) and the dotted
lines (blue lines) indicate that of the polar direction and the equatorial
direction, respectively. 
The survival probability of the equatorial direction is almost 0 in all the panels.
Here we focus on the survival probability of the polar direction.
First, in the panel (a),
the low-energy side of the survival probability becomes finite.
Next, in the panel (b),
the survival probability becomes finite in all the considered energy range.
Finally, in the panel (c), 
the high-energy side of the polar direction only remains finite.
The region of the finite survival probability shifts from
the low-energy side to the high-energy side with time.
 The finite value is close to $\bar{P}\approx 0.7$ that is close 
 to the case of the complete non-adiabatic state.
It is noted that the survival probability for $\bar{\nu}_e$ 
can be given as follows (e.g., \cite{dighe}),
\begin{eqnarray}
\bar{P} & = & 
P_{H} \cos ^2 \theta_{12} \cos^2 \theta_{13}+
       (1-P_{H}) \sin^2 \theta_{13},
\end{eqnarray}
 and with the employed value of $\cos^2\theta_{12}=0.7$,  
\begin{eqnarray}
\bar{P} = 0.7 P_{H} \cos^2 \theta_{13} + 
 (1-P_{H}) \sin^2 \theta_{13},
\label{eq:0.7P}
\end{eqnarray}
where $P_{H}$ is the survival probability at the H-resonance. With 
the shock at the H-resonance, $P_{H}$ is close to unity, which makes $\bar{P}$ close 
 to 0.7.

The behaviors in Figure 6 can be understood by
 comparing the resonance point with the shock position.
When the shock front reaches the region of H-resonance (see the region
enclosed by the blue lines in Figure \ref{fig:radH}), 
the steep decline of the density  at the shock front changes the
resonance into non-adiabatic following Eq. (\ref{eq:gamma}).
As a result, 
the survival probability  starts to become finite.
As the shock propagate from the high density region to low energy region, 
the survival probability becomes finite from low-energy side to
high-energy side,
since the density at the resonance point, $\rho_{\mathrm{res}}$, is proportional to
$E_{\nu}^{-1}$ (see Eq. (\ref{eq:rho-res})).
In the equatorial direction, on the other hand, the survival
probability does not change
 (blue dotted lines of Figure \ref{fig:surveb}). 
This is simply because the density near the resonance along the equatorial plane 
hardly changes as already shown in Figure \ref{fig:radE}. 

\FIGURE[h]{
\epsfig{file=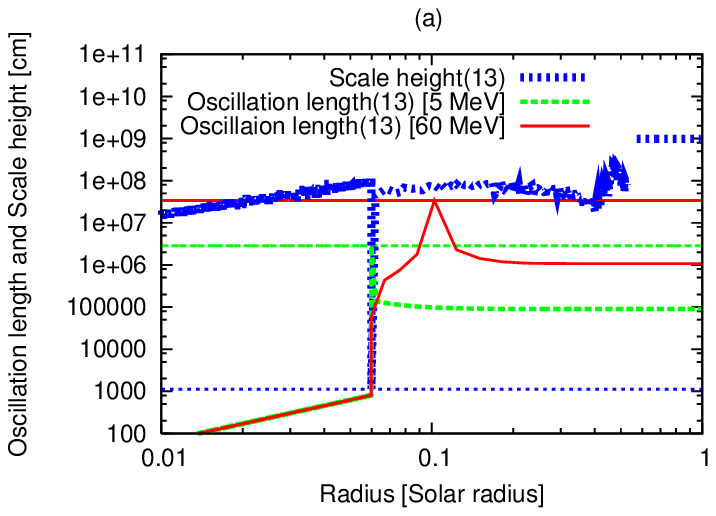,width=0.55\textwidth}
\epsfig{file=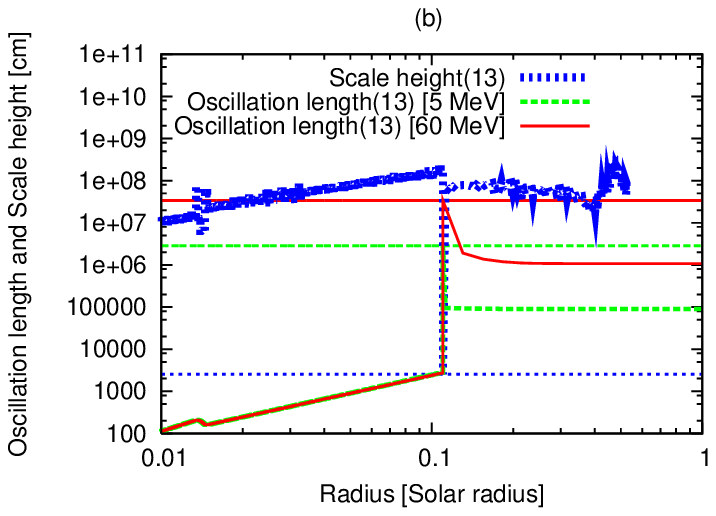,width=0.55\textwidth}
\epsfig{file=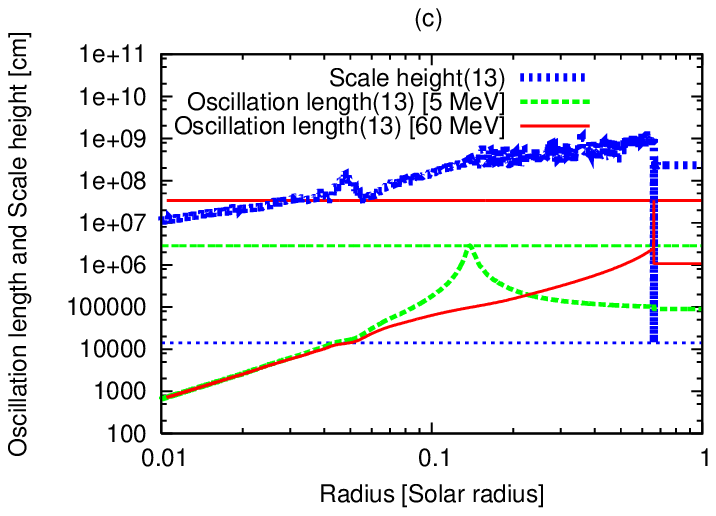,width=0.55\textwidth}
 \caption{\label{fig:osleng13}The dotted blue lines represent the
 scale height (e.g., Eq.(3.3)). The dashed green lines and the solid red lines 
are the oscillation length (Eq. (3.2)) of anti-neutrinos 
for 5 MeV and 60 MeV, respectively. 
Each horizontal lines are the maximum of oscillation length and
 the minimum values of scale height.
It is noted that the label of (13) indicates that the value
 depends sensitively on $\theta_{13}$ and $\Delta m^2_{13}$.
}}


The adiabaticity of the conversion can be evaluated by
 comparing the oscillation length, $L_{\mathrm{osc}}$, with the scale height of
 electron number density, $\delta r$ as follows,
\begin{eqnarray}
\gamma&=&\frac{\delta r}{L_{\mathrm{osc}}}\gg 1\Longleftrightarrow
\delta r \gg L_{\mathrm{osc}}, \label{eq:adiabaticcond}\\
L_{\mathrm{osc}} &\equiv& 
\frac{2 E_{\nu}}{\Delta m^2}
\left[ \left( \frac{2 \sqrt{2}G_F E_{\nu} n_e} 
{\Delta m^2} - \cos 2 \theta \right) ^2 + \sin ^2 2 \theta \right]
^{-\frac{1}{2}},\\
\delta r &\equiv& \frac{\sin 2 \theta}{\cos 2 \theta}
\frac{n_e}{| d n_e/ d r |},
\end{eqnarray}
When the scale height is shorter than the oscillation length,
the resonance is non-adiabatic.
Three panels in Figure \ref{fig:osleng13} show the scale height and the 
oscillation length of neutrinos as a function of radius. 
These panels correspond to the case of (a), (b)
and (c) from top to bottom. 
The dashed lines (green lines) and the solid lines
 (red lines) are the oscillation length of anti-neutrinos
which energy are 5 MeV and 60 MeV, respectively.
Dotted lines (blue lines) are the scale height. Most steepest decline 
 of the scale height shown  in each panel coincides with the position of the 
 shock front. 
Each horizontal line indicates the maximum value of the oscillation length and
the minimum value of the scale height.
 
In Figure \ref{fig:osleng13}(a), the scale height is shown to be 
 shorter than the oscillation length of 5 MeV at the shock front 
at $\sim 0.06 R_{\odot}$ with $R_{\odot}$ being the solar radius (compare 
 the minimum of the blue line with the maximum of the green line, which 
 is shown by the horizontal green line).
Therefore, the H-resonance of 5 MeV becomes non-adiabatic 
(the survival probability $\sim$ 0.7). On the other hand, 
 the scale height is larger, albeit slightly,  than the oscillation length of 
60 MeV, making the resonance adiabatic (the survival probability $\sim$ 0).
In Figure \ref{fig:osleng13}(b), so the scale height is smaller than the
oscillation length of 5 MeV and 60 MeV at the shock front ($\sim 0.1 R_{\odot}$),
 the H-resonance for almost all the neutrino energies become non-adiabatic, making the
survival probability close to be 0.7. Figure \ref{fig:osleng13}(c) is a 
direct opposition as for the resonance conditions to the case of (a).
The oscillation length for the high-energy neutrinos is longer than for the 
 low-energy neutrinos (Eq. (3.2)). This is the reason of the slight increase 
 in the survival probability for the high-energy side seen
 in the equatorial direction (Figure \ref{fig:surveb}).

\subsection{Neutrino spectra of $\bar{\nu}_e$}

\FIGURE[t]{
\epsfig{file=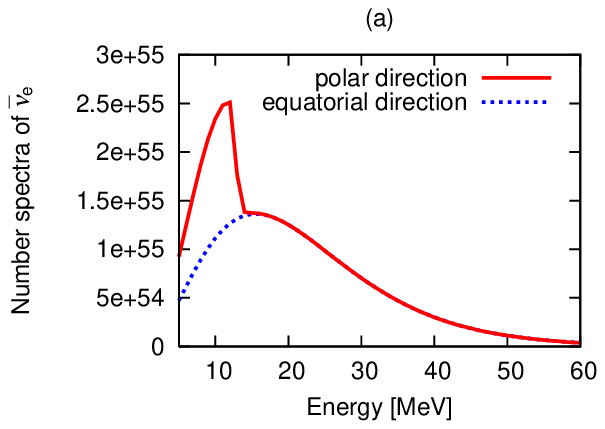,width=0.55\textwidth}
\epsfig{file=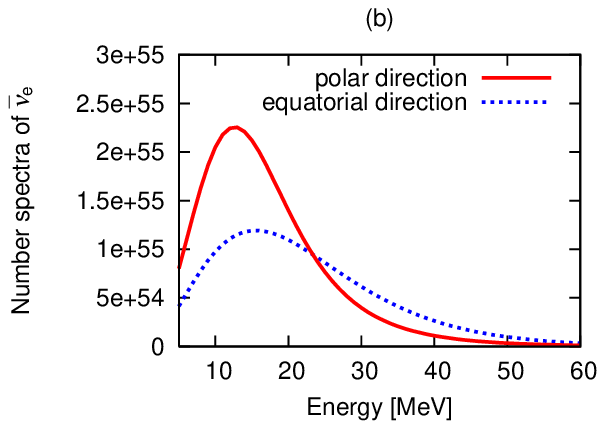,width=0.55\textwidth}
\epsfig{file=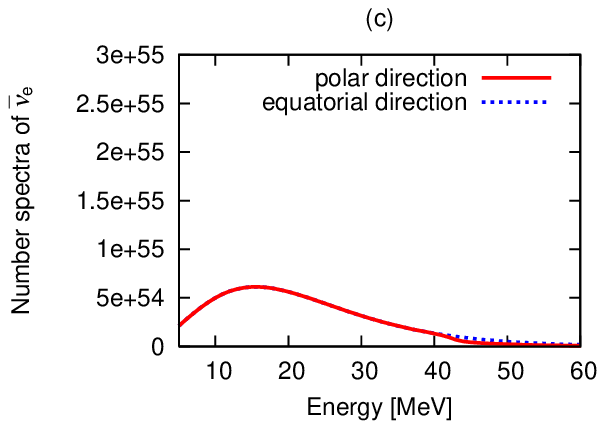,width=0.55\textwidth}
 \caption{\label{fig:spctreb}$\bar{\nu}_e$ spectra at the surface of the star
  in the case of (a), (b) and (c) from top to bottom.  
  We assume the inverted mass hierarchy and $\sin ^2 2
 \theta_{13}=10^{-3}$. The solid lines (red lines) and the dotted
  lines (blue lines) are the spectra of polar direction and equatorial
 direction, respectively. }}


Now we move on to estimate the neutrino spectra at the surface of the star using 
 the calculated survival probability in the previous section, 
 and the original neutrino spectra in section 2.2. 

From Eq. (\ref{eq:spctra}), the spectra of $\bar{\nu}_e$ from the star
 ($\phi ^{\mathrm{SN}}_{\bar{\nu}_e}$) is expressed as,
\begin{eqnarray}
\phi ^{\mathrm{SN}}_{\bar{\nu}_e}
   & = & \bar{P}\phi^{0}_{\bar{\nu}_e}+
        (1-\bar{P})\phi^{0}_{\nu_x} \nonumber \\
   & = & \bar{P}(\phi^{0}_{\bar{\nu}_e} - \phi^{0}_{\nu_x}) +\phi^{0}_{\bar{\nu}_e}.
\label{spec_ref}
\end{eqnarray}
The original spectra of the low-energy side are 
$\phi ^{0}_{\bar{\nu}_e} > \phi ^{0}_{\nu_x}$ 
(see Figure \ref{fig:originalspect}), thus the sign of
$(\phi ^{0}_{\bar{\nu}_e}-\phi ^{0}_{\nu_x})$ 
 becomes positive.
Therefore, $\phi ^{\mathrm{SN}}_{\bar{\nu}_e}$ increases compared with
 the no-shock case because $\bar{P}$ can be finite due to the shock-passage
 (e.g., section 3.1).
On the other hand, the original spectra of the high-energy side are
$\phi ^{0}_{\bar{\nu}_e} < \phi ^{0}_{\nu_x}$,
making the sign of 
($\phi ^{0}_{\bar{\nu}_e} - \phi ^{0}_{\nu_x}$) negative, 
leading to the decrease in $\phi ^{\mathrm{SN}}_{\bar{\nu}_e}$ 
 in comparison with the no-shock case. 
It should be noted that the energy satisfying 
$\phi ^{0}_{\bar{\nu}_e}=\phi ^{0}_{\nu_x} $ is
about 23.3 MeV (See Figure \ref{fig:originalspect}). 

Figure \ref{fig:spctreb} is the energy spectra of $\bar{\nu}_e$ 
in the case of (a), (b) and (c) from top to bottom.
The solid lines (red lines) and the  dotted lines (blue lines) are
the spectra of the polar direction and the equatorial direction,
respectively.
At the panel (a), an enhancement of the low-energy side of the polar direction
 is seen. This is because the survival probability at the low energy side is finite
as shown in the panel (a) of Figure \ref{fig:surveb}. 
The effect of the shock on the neutrino spectra 
 also shifts from the low-energy to the high-energy side as the shock front goes 
outward. In fact, the energy spectrum of the polar direction becomes softer 
in the high-energy side than for the equatorial direction 
 as shown in Figure \ref{fig:spctreb} (b).

\subsection{Expected event number of $\bar{\nu}_e$ at the Super-Kamiokande 
 detector}\label{sec:event-rate-nu_e_bar}

From Eq. (2.6), we calculate the expected event number at SK for a 
 Galactic supernova.
 Figure \ref{fig:obseb} shows the time evolution of the event number,
 in which the solid line (red line) and the dotted line (blue line) are 
 for the polar and the equatorial direction, respectively. 
 The shock effect is clearly seen.
The event number of the polar direction shows a steep decrease,
 marking the shock passage to the H-resonance layer.
Moreover, we can see a slight enhancement for the polar direction around 0.5
sec, because of the increase of the low-energy neutrinos by the shock
propagation that is seen in the panels (a) and (b) of Figure
\ref{fig:spctreb}.

It is noted that the decrease in the events comes mainly 
from the decrease of the high-energy neutrinos rather than 
the increase of the low-energy neutrinos. This is because the cross section of 
 $\bar{\nu}_e + p \rightarrow e^+ + n$, main reaction for detection, is 
 proportional to the square of the neutrino energy ($E_{\nu}^2$).
 The change of the event number by the shock passage 
 is about 36\% of the event number without the shock.
 Since the expected events are $\sim 2500$ at the sudden decrease 
(Figure \ref{fig:obseb}), it seems to be quite possible to identify such a feature
 by the SK class detectors. 
 Such a large number imprinting the shock effect, 
 possibly up to two-orders-of magnitudes larger than the ones 
 predicted in the neutrino-driven explosions (e.g., Fig 11 in \cite{Tomas}),
 is thanks to the mentioned early shock-arrival to the resonance layer, 
 peculiar for magnetic explosions.

\DOUBLEFIGURE[t]
  {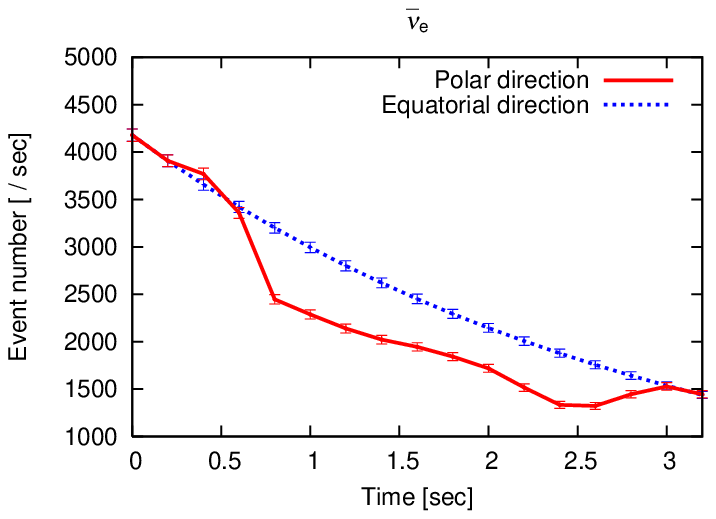}{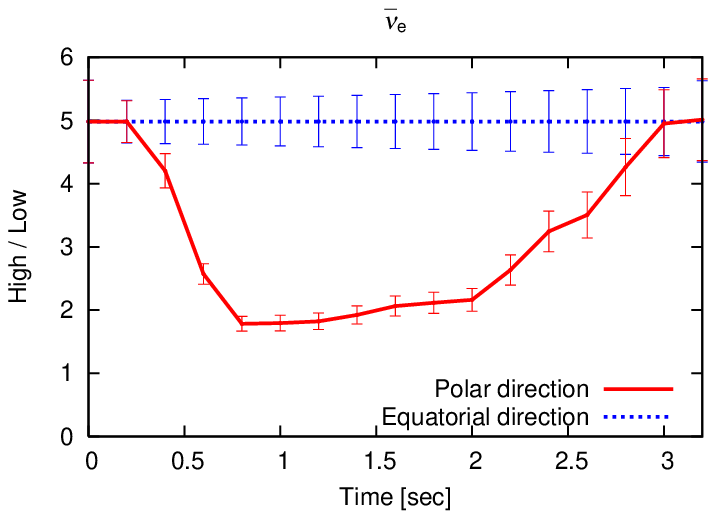}
  {\label{fig:obseb}The expected event number of $\bar{\nu}_e$ in the
  SK as a function of the time.
  The solid line (red line) and the the dotted line (blue line) is 
 for the polar and the equatorial direction, respectively.
  The supernova is assumed to be located at the distance of 10 kpc. 
The error bars represent the 1$\sigma$ statistical errors only.
}
 {\label{fig:ratioeb}
Same as Figure 9, but for the ratio of high- to low-energy events
 defined in Eq. (3.7).}

Here we define a ratio of the 
high-energy to the low-energy event numbers as in \cite{Takahashi1}: 
\begin{eqnarray}
\mathrm{R(H/L)} = \frac{\mathrm{event} \,\, \mathrm{number}
          \,\, \mathrm{of}
          \,\,  20 < E < 60 \,\, \mathrm{MeV}}
{\mathrm{event} \,\, \mathrm{number} \,\, \mathrm{of}
          \,\,  5 < E < 20 \,\, \mathrm{MeV}} .
\label{eq:ratio}
\end{eqnarray}
It is shown in Figure \ref{fig:ratioeb} 
that the ratio could be a helpful tool to identify
 the shock effect more clearly.

\subsection{Expected event number of $\nu_e$ at the Super-Kamiokande 
detector}

Now we proceed to discuss the shock effect on the L-resonance (Figure \ref{fig:radL}).

In the L-resonance region, the shock influences not $\bar{\nu}_e$ 
but $\nu_e$ in the inverted mass hierarchy.
We calculate the survival probability of $\nu_e$, $P(E_{\nu})$, 
in the same manner
described in Section \ref{sn-neu-model}.
As will be shown below, the qualitative behavior of $\nu_e$ in the 
 L-resonance closely resembles to that of $\bar{\nu}_e$ in the H-resonance.
Just like the case of the H-resonance (section \ref{Surv-prob-H}), 
the survival probability of $\nu_e$ is about 0.3 in absence of the 
  shock, becomes 0.7, when the shock 
reaches to the L-resonance layer.
Also as in the case of $\bar{\nu}_e$, the shock effect on the survival probabilities 
shifts from the low- to high- energy side as the shock propagates outward, 
leading to the change in the observed spectra of $\nu_e$ (Figure 11).

\FIGURE[t]{
\epsfig{file=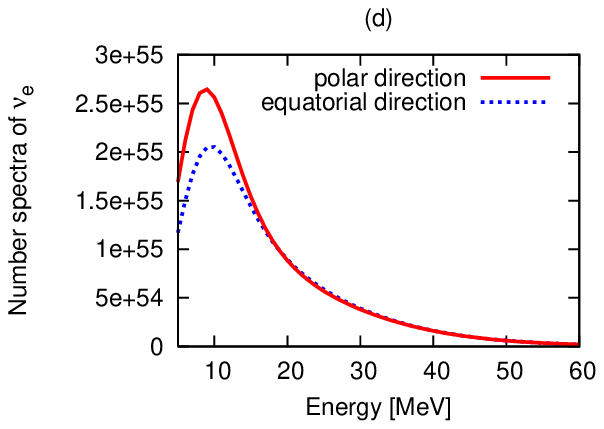,width=0.55\textwidth}
\epsfig{file=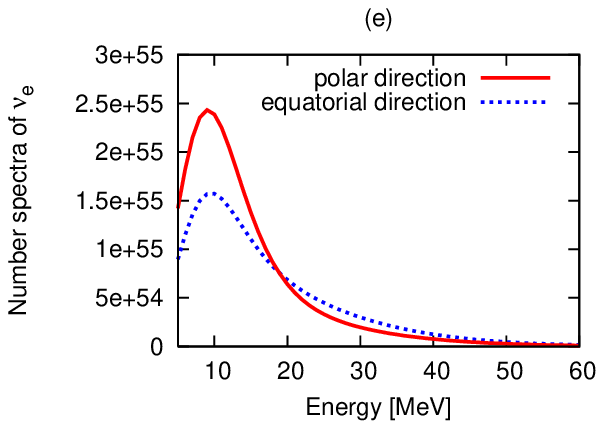,width=0.55\textwidth}
\epsfig{file=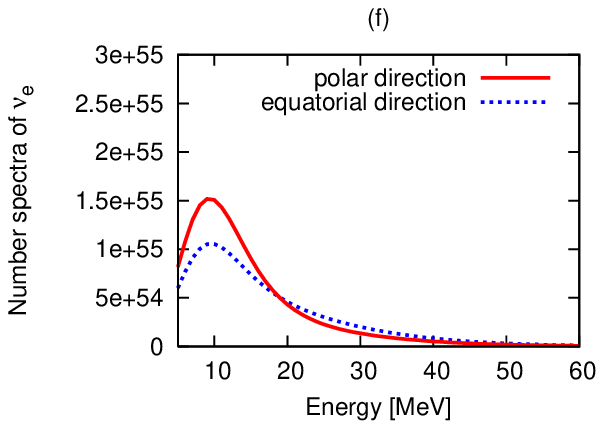,width=0.55\textwidth}
 \caption{\label{fig:spctre}The $\nu_e$ spectra in the case of (d), (e)
 and (f) from top to bottom. We assume the  
  inverted mass hierarchy and $\sin ^2 2 \theta_{13}=10^{-3}$.}}


With these survival probabilities, we calculate the expected 
event number at the SK for a Galactic source (Figure \ref{fig:obse}).
It can be seen that the event number of the polar direction initially increases 
and then decreases compared with the one for the equatorial direction, which is 
 due to the increase of the low-energy neutrinos and due to decrease 
 of the high-energy neutrinos, respectively. 
The event number of $\nu_e$ become much fewer than that of $\bar{\nu}_e$.
This is because the main reaction for detecting $\nu_e$ is 
$\nu_e + e^- \rightarrow \nu_e + e^-$, and the cross section is  
 about $10^{-2}$ times smaller than that of $\bar{\nu}_e$.
Note that the cross section of $\nu_e + e^- \rightarrow \nu_e + e^-$ is 
largest among neutrino-electron scattering reactions (e.g.,
 \cite{hirata1988}) 
 and that this forward scattering reaction could be an important tool 
 for identifying the direction to the supernova 
(\cite{nakahata, burrows1992, beacom1999, tomas2003}). 
As in the case of the $\bar{\nu}_e$,
R(H/L) defined in Eq.(\ref{eq:ratio}) could be a good probe, especially
 when the event number is smaller here (Figure \ref{fig:ratioe}).
 The ratio begins to decrease near around $1$ s, which might be an 
 observable signature of the shock entering to the L-resonance.
While the statistical errors are eye-catching compared to those of the 
H-resonance due to the small event numbers here (compare Figure \ref{fig:obseb}), 
the overlap of the error bars from 1.5 to 2.5s in the high/low-ratio is 
fortunately small, which could mark a marginally distinguishable
feature.

To our knowledge, the possibility of observing the shock feature in the L-resonance 
 has been never claimed before. This is because in the conventional delayed 
 explosion models of core-collapse supernovae, the neutrino oscillation
 in L-resonance is thought to be minor
because the original neutrino luminosity decreases to be so small
until the shock wave reaches to the L-resonance.
Due to the small cross section of $\nu_e$, 
the detection of the shock effect should not be easy even for the case of 
magnetic explosions considered here. However we speculate the detection to be 
more promising when the proposed next-generation megaton-class detectors
 such as the Hyper-Kamiokande ($\sim$0.5-Mton) \cite{HK} and the Deep-TITAND 
($\sim$5-Mton) \cite{titand} 
are on-line, possibly making the event number up to two-orders-of-magnitudes higher
 than for the SK.

\DOUBLEFIGURE[t]
  {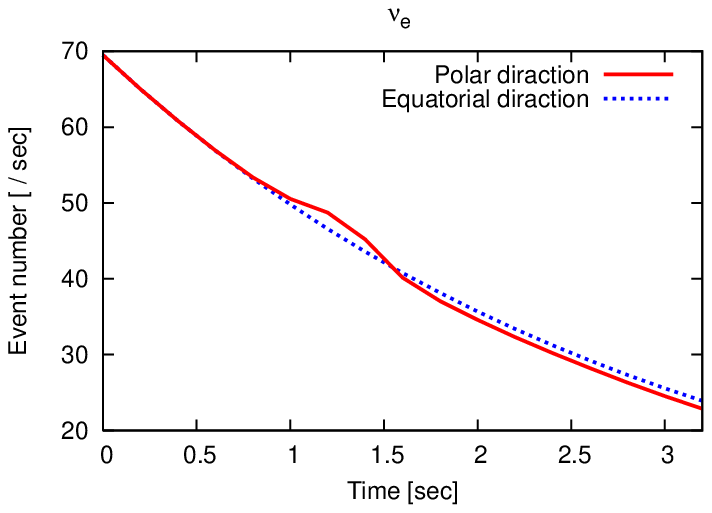}{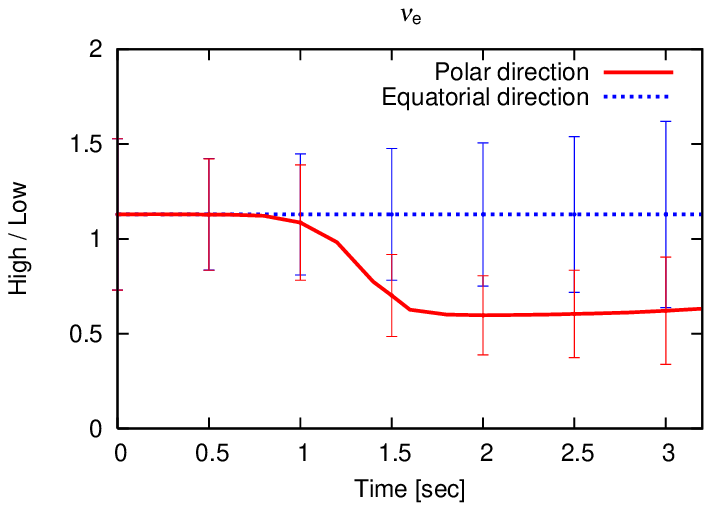}
  {\label{fig:obse}The expected event number of $\nu_e$ in the
  SK as a function of time. The solid line (red line) and
  the dotted line (blue line) for the polar and equatorial direction, respectively.
  The supernova is assumed to be located at the distance of 10 kpc.}
{\label{fig:ratioe} Same as Figure 12, but for the ratio defined in Eq. (3.7).
The error bars represent the statistical errors only.
}


\sout{
We used the Type-Ic supernova model in this study.
This is thought to be an explosion of the star without the envelope of
hydrogen layer and helium layer.
When the shock wave reaches L-resonance, the density gradient in
L-resonance becomes sharp.
If the star has the envelope, the density gradient in L-resonance
returns to be gradual after the shock wave passed L-resonance.
However, if the star does not have the envelope, like this paper,
the neutrinos become in the state of non-adiabatic at long time after
the shock wave passing L-resonance.}

\section{Discussion}
\subsection{Dependence of $\theta_{13}$ on neutrinos}\label{dis-theta-13}

\FIGURE[t]{
\epsfig{file=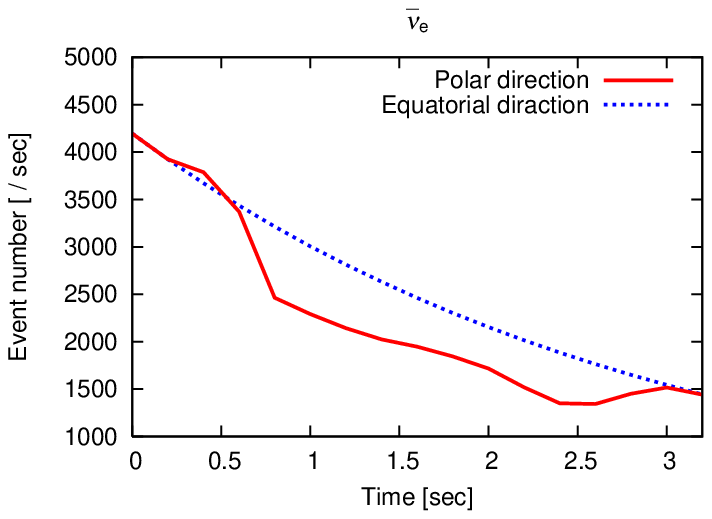,width=0.45\textwidth}
\epsfig{file=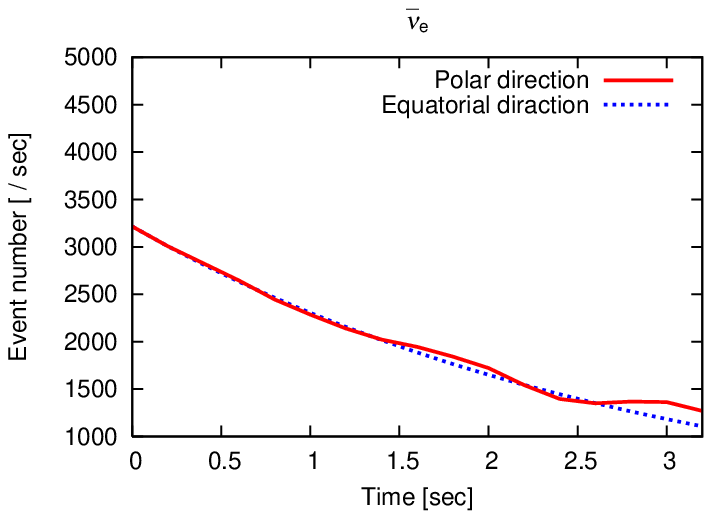,width=0.45\textwidth}
 \caption{\label{fig:obsparameter}
The expected event number of  $\bar{\nu}_e$ are shown.
The expected event number in the left panel are calculated with $\sin ^2 2
 \theta _{13}=10 ^{-2}$ and
 that in the right panel are calculated with $\sin ^2 2\theta _{13}=10^{-5}$.
   The solid lines (red lines) and the dotted
  lines (blue lines) are the event number in the polar direction and
 the equatorial direction, respectively.}}


First of all, we discuss the dependence of $\theta_{13}$ on the expected
event number. 
Considering the uncertainty of $\theta_{13}$,
we calculate the event number changing $\sin ^2 2\theta_{13}=10^{-2}$
and $10^{-5}$ 
in addition to the fiducial value of $10^{-3}$.

The two panels of Figure \ref{fig:obsparameter} show
 the event number of $\bar{\nu}_e$ for $\sin^2 2 \theta_{13}=10^{-2}$ (left) 
and $10^{-5}$ (right).
 It can be seen that the event number of $\sin ^2 2 \theta_{13}=10^{-2}$ 
is larger than that of $\sin ^2 2 \theta_{13}=10^{-5}$, 
regardless of the viewing direction.
Moreover, the event number for $\sin ^2 2 \theta_{13} =10^{-2}$ in 
 the polar direction shows a much steeper shock effect than 
for $\sin ^2 2 \theta_{13}=10^{-5}$. Those features can be understood as
follows.

As already mentioned, the change in the event number is sensitive to the 
 adiabaticity condition (e.g., Eq. (\ref{eq:gamma}).
If $\theta _{13}$ is relatively large 
(e.g. $\sin ^2 2 \theta_{13}=10^{-2}, 10^{-3}$) and there is no shock
wave in the resonance, the adiabatic condition is satisfied for the H-resonance,
 converting almost all  $\bar{\nu}_x$ to $\bar{\nu}_e$.
Since the average energies of neutrinos in the supernova 
 environment are in the following order,
$\bar{E}_{\nu_e} < \bar{E}_{\bar{\nu}_e} < \bar{E}_{\nu_x,
\bar{\nu}_x}$,
 the average energy of $\bar{\nu}_e$ becomes higher than the original $\bar{\nu}_e$
 due to the flavor conversion from $\bar{\nu}_x$.
 On the other hand, if $\theta_{13}$ is small (e.g., $10^{-5}$) or there is a shock 
 wave in the resonance,  the adiabatic condition is not satisfied.
In this non-adiabatic case in the H-resonance, the original $\bar{\nu}_e$ spectrum almost keeps unchanged.

As stated in section \ref{sec:event-rate-nu_e_bar},
the cross section for detecting $\bar{\nu}_e$ depends on $E_{\nu}^2$.
Therefore the $\bar{\nu}_e$ event number of the adiabatic case becomes larger than 
that of the non-adiabatic case.
If the resonance is closely non-adiabatic always, the shock effect becomes 
 very minor, which is the case for the right panel in Figure \ref{fig:obsparameter}. 
We also calculate R(H/L) defined by Eq.(\ref{eq:ratio}). 
In the case of $\sin ^2 2 \theta_{13}=10^{-2}$,
R(H/L) of $\bar{\nu}_e$ is similar to Figure \ref{fig:ratioeb}.
Conversely, the time evolution of R(H/L) of $\bar{\nu}_e$ of 
$\sin ^2 2 \theta_{13}=10^{-5}$
does not show the shock feature.

\subsection{Difference of the original neutrino spectrum}\label{dif-spe}

\DOUBLEFIGURE[t]
  {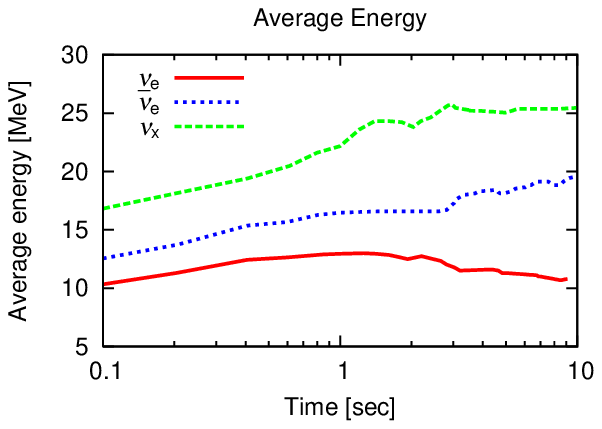}{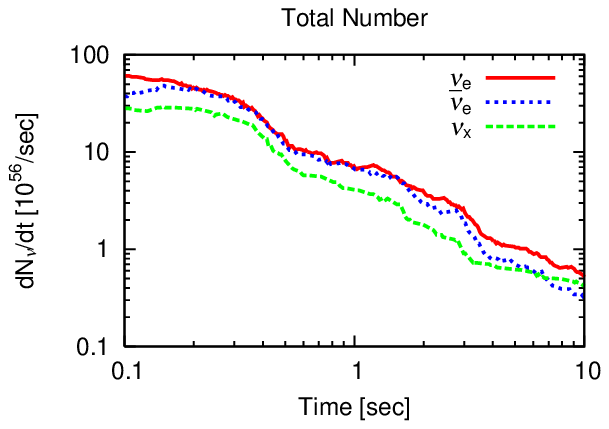}
  {\label{fig:eveene}
 The evolution of average neutrino energies for the
 time-dependent spectra model
 \cite{Fogli2} (see text for more details).  
The solid line (red line), dashed line (green line) and the dotted line 
(blue line) represents the evolution of the average energy of 
$\nu_e$, $\bar{\nu}_e$ and 
$\nu_x$, respectively.}
{\label{fig:totalnum} 
 Same as Figure 15, but for the neutrino emission rate
as a function of time in Eq. (4.1) (e.g., \cite{Fogli2}). }

\FIGURE[t]{
\epsfig{file=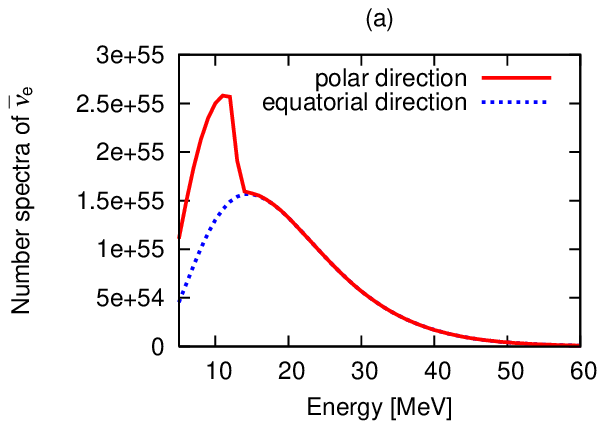,width=0.45\textwidth}
\epsfig{file=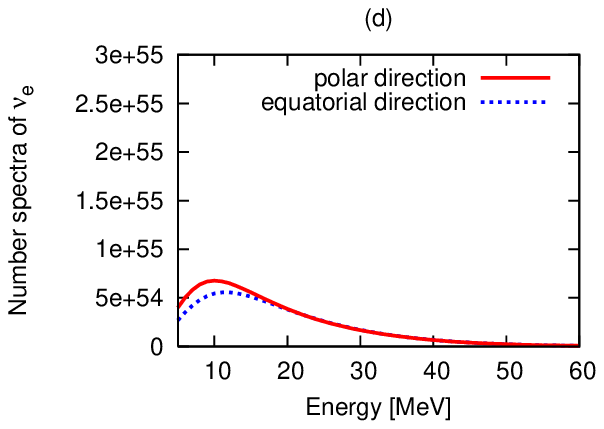,width=0.45\textwidth}
\epsfig{file=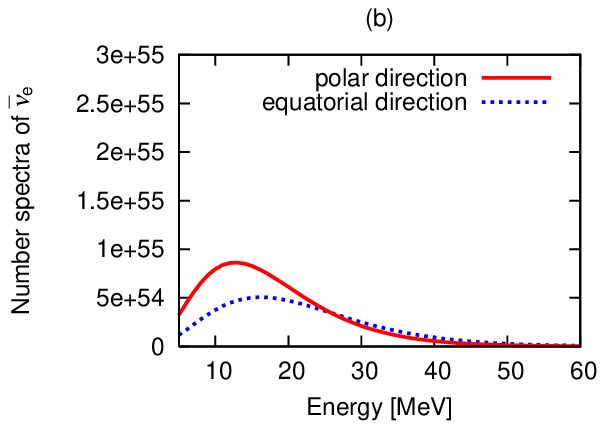,width=0.45\textwidth}
\epsfig{file=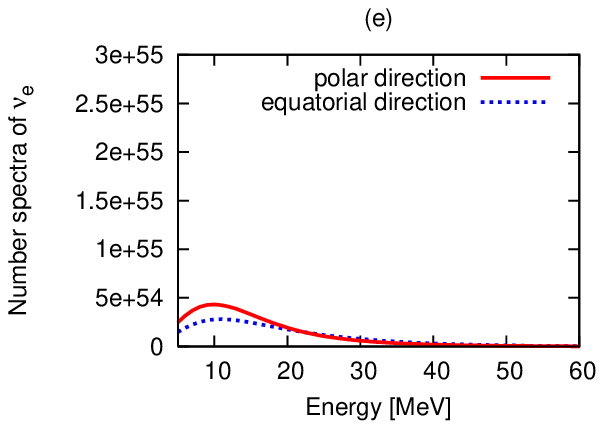,width=0.45\textwidth}
\epsfig{file=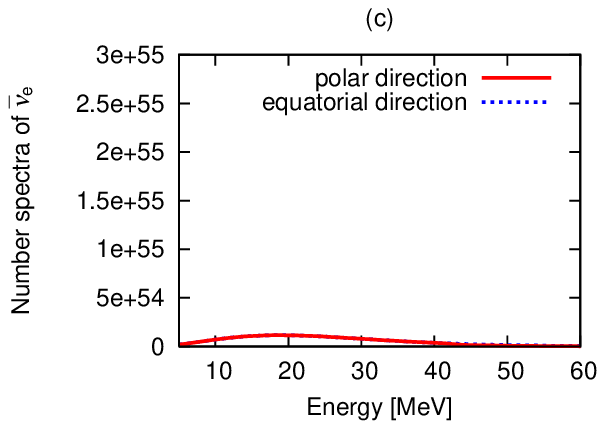,width=0.45\textwidth}
\epsfig{file=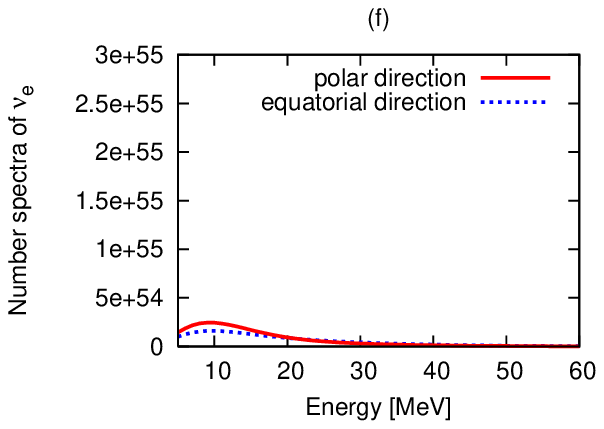,width=0.45\textwidth}
 \caption{\label{fig:anotherspcteb}
The neutrino spectra in the case of (a)$-$(f) using 
the time-dependent original spectra.
Left panels are spectra of
 $\bar{\nu}_e$, and right figures are $\nu_e$.
The solid lines (red lines) and the dotted
  lines (blue lines) are the spectra in the polar direction and
 the equatorial direction, respectively.
We assume $\sin ^2 2 \theta _{13}=10 ^{-3}$ and inverted mass
 hierarchy.} }

In this section, we discuss the variations of the original neutrino
spectra, 
 and its effect on the obtained results.
As already mentioned in section \ref{sn-neu-model},
 we have employed the one, which is assumed to decay exponentially with 
 a timescale of $t_{\rm L} =$ 3 s 
with the average neutrino energies kept unchanged and is 
 adjusted to be consistent with the total neutrino energies in the model of 
Lawrence Livermore group \cite{totani-sato-del}. Such a treatment
 has been often employed so far in the supernova neutrino and nucleosynthesis 
studies (e.g. Yoshida et al. 2006 \cite{yoshida2} and reference therein).
 When we set $t_{\rm L}$ to be 6 sec, the obtained results are qualitatively 
unchanged though the event number becomes smaller up to a factor of $\sim$ 2 than for the decay-time of 3 s.

To explore the uncertainty of the original spectra, 
 we here employ another original neutrino spectra used in \cite{Fogli2} as follows,
\begin{eqnarray}
\phi^0_{\nu} = \frac{d N_{\nu}}{dt}\phi(E_{\nu}),
\end{eqnarray}
where $dN_{\nu}/dt$ is a neutrino emission rate, 
and $\phi(E)$ is a normalized energy spectra defined as, 
\begin{eqnarray}
\phi (E_{\nu}) = \frac{ (\alpha + 1)^{\alpha + 1}}{\Gamma (\alpha + 1)}
          \left( \frac{E_{\nu}}{\langle E_{\nu} \rangle} \right) ^{\alpha }
          \frac{e ^{-(\alpha + 1)E_{\nu}/ \langle E_{\nu} \rangle}}
           {\langle E_{\nu} \rangle}.
\label{eq:phi}
\end{eqnarray}
$\langle E \rangle $, $\alpha$, and $\Gamma$ is the average energy, an
energy shape parameter, and the gamma function, respectively \cite{Fogli2}. 
For simplicity, we have taken $\alpha=3$ for all flavors \cite{Tomas}.
In this model, $\frac{d N_{\nu}}{dt}$ and  $\langle E \rangle $ follow the calculation of
 the Lawrence Livermore group \cite{totani-sato-del}.
The time evolution of the average energy and the neutrino emission rate 
 is shown in Figure \ref{fig:eveene} and \ref{fig:totalnum}, respectively.
The solid line (red line), the dashed line (green line) and the dotted line 
(blue line)  corresponding that of $\nu_e$, $\bar{\nu}_e$ and 
$\nu_x$, respectively.
Hereafter we call this spectra as ``time-dependent'' original spectra.

Figure \ref{fig:anotherspcteb} shows the neutrino spectra at the SK calculated 
 by the mentioned procedure in section \ref{sec:event-rate-nu_e_bar}, in which 
 left and light panel is
 for $\bar{\nu}_e$ and $\nu_e$, respectively. 
 It can be shown that the neutrino spectra are qualitatively 
insensitive to the difference 
in the original neutrino spectra examined here. 
However significant difference appears in the event number.
Figure \ref{fig:anotherobseb} and \ref{fig:anotherobse} is 
the event number of $\bar{\nu}_e$ and $\nu_e$, respectively.
The difference of the event number between the polar and the equatorial direction 
becomes much smaller than that of our fiducial model shown in Figures \ref{fig:obseb}
 and \ref{fig:obse}.
This is because the neutrino flux of the time-dependent spectra decreases
 already before the shock reaches to the resonance layers. 

R(H/L) is shown in Figures \ref{fig:anotherratioeb}
 (for $\bar{\nu}_e$) and \ref{fig:anotherratioe} (for $\nu_e$).
In the equatorial direction of Figure \ref{fig:anotherratioeb}, 
the ratio gradually increases although it is constant 
in Figure \ref{fig:ratioeb}.
This reflects the increase of the average energy of $\bar{\nu}_e$ 
with time (see the dotted line of Figure \ref{fig:eveene}).
In contrast, the ratio of the equatorial direction is almost flat in
 Figure \ref{fig:anotherratioe}, reflecting the evolution of the
 average energy of $\nu_e$. 
The error bars here are shown to be larger than for Figure \ref{fig:ratioe},
 simply because the event number of $\nu_e$ here becomes smaller than for the 
 original neutrino spectrum in section 2.2 (compare Figure 12 with 19).
Albeit sensitive to the feature of the 
 original spectra, it is found that the ratio for the polar direction
  becomes generally smaller than for the equatorial
direction due to the shock effect (Figures 10, 20, 21) for 
 relatively large value of $\theta_{13}$.

  Those results indicate that the shock effect on the event number discussed so far,
 is subject to the large uncertainty of the 
 original neutrino spectra. What is qualitatively new pointed out in this paper is 
 that in the MHD explosions, the timescale of the shock-arrival 
 could be shorter than the decay timescale of the neutrino flux, albeit 
 depending on the evolution of the neutrino spectra.
It should be also mentioned that if the energy difference between $\bar{\nu}_e$ and
 $\nu_x$ becomes small due to detailed neutrino interactions 
near the neutrinosphere 
\cite{raffelt2001, keil2003}, the shock effect should be also small.
 To determine the original neutrino spectra in a consistent manner
 with the shock-dynamics in the supernova envelope requires more sophisticated 
MHD modeling of core-collapse supernova, in which detailed microphysics is also 
 included. Several groups are really 
pursuing it with the use of advanced
numerical techniques \cite{Burrows07, Takiwaki09}, 
 however still unfeasible at present.

In this paper, we neglected the effect of neutrino self-interactions.
Among a number of important effects possibly created by self-interactions
 (e.g.,\cite{duan_rev, duan2006}),  
we choose to consider the effect of single spectral swap 
of $\bar{\nu}_e$ and $\nu_x$ suggested by Fogli et al. \cite{fogli1}.
In our model, the original spectrum was 
$\phi_{\bar{\nu}_e}>\phi_{\nu_x}$ in the low-energy side, and 
$\phi_{\bar{\nu}_e}<\phi_{\nu_x}$ in the high-energy side.
Due to the spectral swap, the above inequality sign reserves. 
Thus the influence of the shock should occur also oppositely
 (see Eq.(\ref{spec_ref}) and discussion in section 3.2). Therefore the
event number of neutrinos in the polar direction is expected to increase
by the influence of the shock.
To draw a robust conclusion to the self-interaction effects on our models naturally 
requires more sophisticated analysis (e.g., \cite{dasgup1}), 
which we pose as a next step of this study.

As a prelude to more realistic computations of the flavor conversions
 taking into account the effects of neutrino self-interaction,
also with more realistic input of the neutrino 
 spectra, our findings obtained
 in a very idealized modeling should be the very first step towards the 
  prediction of the neutrino signals from magnetic explosions
 of massive stars.

\DOUBLEFIGURE[t]
  {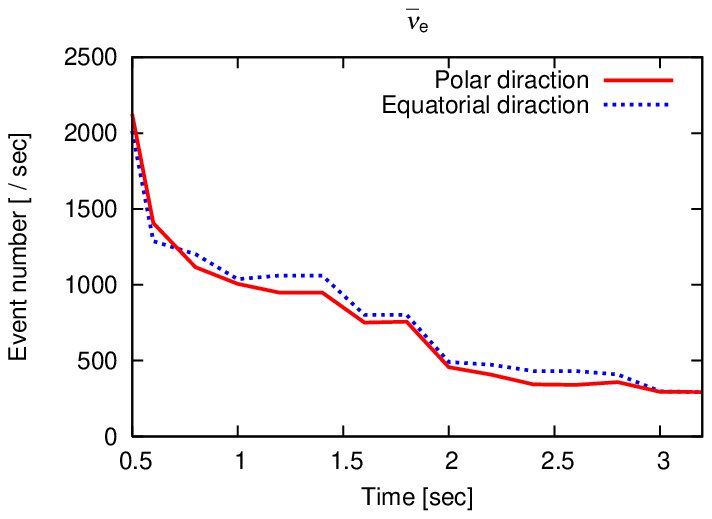}{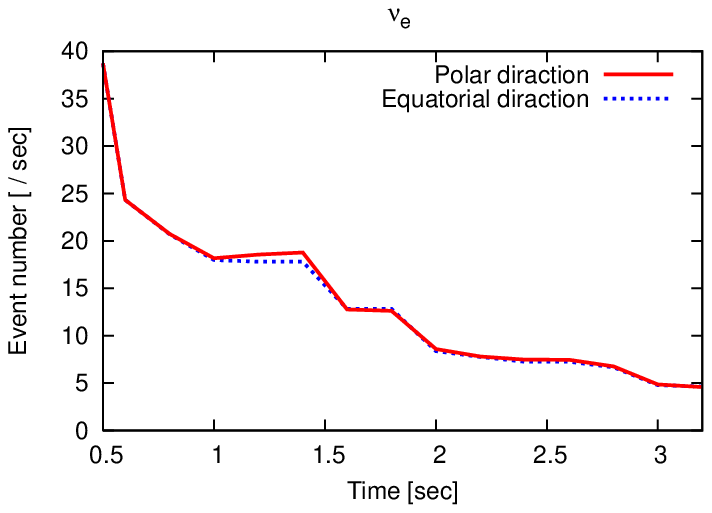}
  {\label{fig:anotherobseb}The expected event number of
 $\bar{\nu}_e$ using  
 the time-dependent original spectra.
 The solid lines (red lines) and the dotted
  lines (blue lines) are the event number for the polar and
 equatorial direction, respectively. The source is assumed to be at $10$ kpc.}
  {\label{fig:anotherobse} Same as Figure 18 but for 
 $\nu_e$.}
\DOUBLEFIGURE[t]
{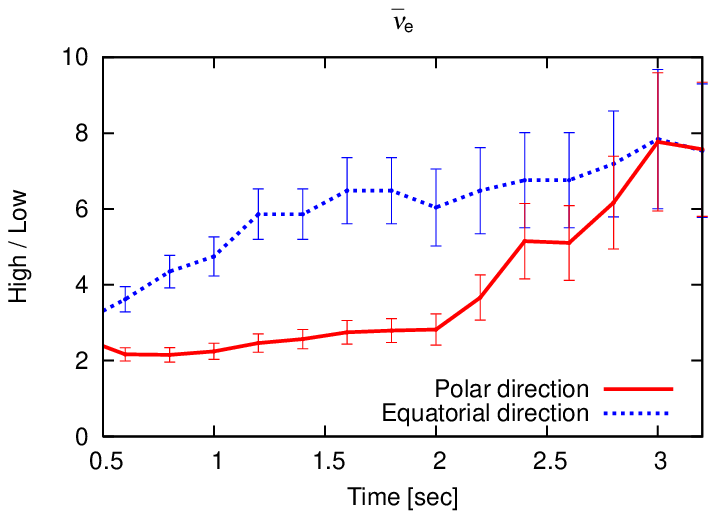}{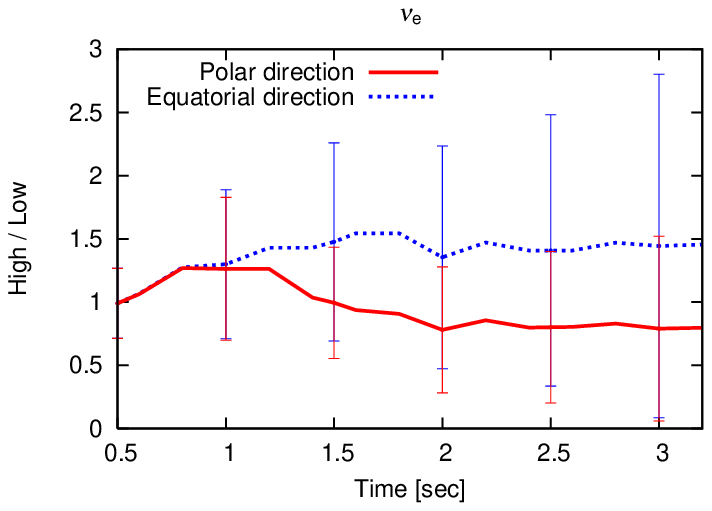}
{\label{fig:anotherratioeb}The ratio of event number of $\bar{\nu}_e$
  using the time-dependent original spectra.
  The solid lines (red lines) and the dotted
  lines (blue lines) are the event number in polar direction and
 equatorial direction, respectively.} 
{\label{fig:anotherratioe} Same as Figure 20 but for $\nu_e$.}

\section{Summary}

We studied neutrino oscillations from core-collapse supernovae that produce 
magnetohydrodynamic (MHD) explosions, which are attracting great attention 
recently as a possible relevance to magnetars and/or long-duration gamma-ray bursts.
 Based on a recent supernova simulation producing MHD explosions till the 
 shock break-out, we calculated numerically 
the flavor conversion in the highly non-spherical envelope 
through the pure matter-driven MSW effect. As for the neutrino spectra 
at the neutrino sphere which the MHD simulations lack, we employed the 
 two variations based on the Lawrence Livermore simulation, with or without the 
 evolution of the average neutrino energies. 
 With these computations, we investigated how the 
explosion anisotropy could have impacts on the emergent 
neutrino spectra and also on the observed neutrino event number at the Super-Kamiokande
 detector for a Galactic supernova. 

 In the case of the inverted mass hierarchy with 
 a relatively large $\theta_{13}$ 
($\sin^2 2 \theta_{13}\gtrsim 10^{-3}$),
we demonstrated that survival probabilities of 
$\bar{\nu}_e$ and $\nu_e$ seen along the rotational axis of the MHD supernovae 
(i.e., polar direction), could be significantly different from those seen from the 
equatorial direction.  The event numbers of $\bar{\nu}_e$ 
 observed from the polar direction show steepest decrease, reflecting 
 the passage of the magneto-driven shock to the so-called high-resonance regions. 
 We pointed out that such a shock effect, depending on the original neutrino spectra, 
appears also for the low-resonance regions, which could lead to a noticeable 
decrease in the $\nu_e$ signals.
This reflects a unique nature of the magnetic explosion featuring a very early 
shock-arrival to the resonance regions, which is 
in sharp contrast to the delayed neutrino-driven 
 supernova models in spherical symmetry. Our results suggested that the two 
features in the $\bar{\nu}_e$ and $\nu_e$ signals, if visible to the 
Super-Kamiokande for a Galactic supernova, could be an observational signature 
 of magneto-driven supernovae.

One of the nearest supernovae, associated with the long-duration gamma-ray burst (GRB)
 is, SN1998bw (Type Ic) at the distance of $\sim 14$ Mpc \cite{Galama98}.
 As mentioned earlier, such events are likely to be associated with the 
energetic stellar explosions induced by magnetic mechanisms,
 often referred to as hypernovae (e.g., \cite{tanaka} and references therein).
It is interesting to note that the fraction of SNe Ib/c associated with
 long GRBs is small ($\sim 1\% - 10\%$), however, is appeared to coincide with 
 that of hypernovae. Albeit with large uncertainties of GRB statistics, 
 the link between hypernovae and long GRBs seems to be observationally supported 
(see discussion in \cite{fryer2007} and reference therein).
 Recently the proposals of Mton-class detectors 
 such as Hyper-Kamiokande \cite{HK} and Deep-TITAND \cite{titand}, are becoming a 
real possibility, by which the shock effect studied here could be 
visible out to the Megaparsec distance scales.
 This implies that the shock effect, if observed (much nicer if observed also
 by the electromagnetic and gravitational-wave observations \cite{kotagw,ott2009}), could provide important
 hints to 
 understand the long-veiled central engines of gamma-ray bursts, 
because the shock arrival time to resonance layers should reflect the activities of the 
engine. We hope that this study could 
 give strong momentum to theorists for making more precise predictions of the 
 neutrino signals from magneto-driven supernovae.

\acknowledgments
We are grateful to H. Suzuki and T. Kajino for fruitful discussions.
K.K. thanks to K. Takahashi and S. Ando for helpful exchanges.
T.T. and K.K. express special thanks to K. Sato and S. Yamada 
for continuing encouragements. 
Numerical computations were in part carried on XT4 and 
general common use computer system at the center for Computational Astrophysics, CfCA, the National Astronomical Observatory of Japan.  This
study was supported in part by the Grants-in-Aid for the Scientific Research 
from the Ministry of Education, Science and Culture of Japan (Nos. 19540309 and 20740150).



\begin{thebibliography}{99}
\bibitem{ando_new}
S.~Ando, J.~F.~Beacom and H.~Yuksel,
  {\it Detection of neutrinos from supernovae in nearby galaxies},
 \prl{95}{2005}{171101}  
\bibitem{heger03}
A.~Heger, C.~L.~Fryer, S.~E.~Woosley, N.~Langer and D.~H.~Hartmann,
  {\it How Massive Single Stars End their Life}
  \apj{591}{2003}{288}

\bibitem{hirata1987}
    K. Hirata, Y. Kajita, M. Koshiba, M. Nakahata and Y. Oyama,
    {\it Observation of a Neutrino Burst from the Supernova SN 1987a},
   \prl{58}{1987}{1490} 

\bibitem{bionta1987}
    R. M. Bionta, G. Blewitt, C. B. Bratton, D. Caspere and A. Ciocio,
    {\it Observation of a Neutrino Burst in Coincidence with Supernova
	SN 1987a in the Large Magellanic Cloud},    
    \prl{58}{1987}{1494}

\bibitem{Sato-and-Suzuki}
    K. Sato  and H. Suzuki,
    {\it  Analysis Of Neutrino Burst From The Supernova In Lmc.}
    \prl{58}{1987}{2722}


\bibitem{1988-Suzuki} 
    H. Suzuki and K. Sato,
    {\it  Statistical Analysis Of The Neutrino Burst From Sn1987a},
    \ptp{79}{1988}{725}



\bibitem{raffelt_review}
G.~G.~Raffelt,
  {\it Physics with supernovae}
\npps{110}{2002}{254} 



\bibitem{Totsuka92}
    Y. Totsuka,
    {\it Neutrino astronomy}
\newjournal{Rept. Prog. Phys.}{RPP}{\bf 55}{1992}{377}

\bibitem{Suzuki99}
    KamLAND Collaboration, A. Suzuki et al.,
    {\it  Present Status of KamLAND},
    \npps{77}{1999}{171}

\bibitem{totani-sato-del}
    T. Totani, K. Sato, H. E. Dalhed and J. R. Wilson,
    {\it  Future detection of supernova neutrino burst and explosion
	mechanism},
    \apj{496}{1998}{216}
    [\astroph{9710203}]

\bibitem{kamland-beacom}
    J. F. Beacom, 
   {\it  Supernovae and neutrinos},
    \npps{118}{2003}{307}
    [\astroph{0209136}]

\bibitem{dighe}
    A. S. Dighe and  A. Y. Smirnov,
    {\it Identifying the neutrino mass spectrum from a supernova
	neutrino burst},
    \prd{62}{2000}{033007}
    [\hepph{9907423}].


\bibitem{Fogli4}
    G. Fogli, E. Lisi, D. Montanino and A. Palazzo,
    {\it  Identifying the neutrino mass spectrum from the neutrino burst
	from a supernova.}
    \prd{65}{2002}{073008}
    [\hepph{9907423}].

\bibitem{Lunardini03}
    C. Lunardini and A. Yu. Smirnov
    {\it Probing the neutrino mass hierarchy and the 13-mixing with
	supernovae }
    \newjournal{JCAP}{JCAPA}{0306}{2003}{009}
    [\hepph{0302033}]

\bibitem{Dighe03a}
    A. S. Dighe, M. T. Keil and G. G. Raffelt,
    {\it  Detecting the neutrino mass hierarchy with a supernova at
IceCube}
    \newjournal{JCAP}{JCAPA}{0306}{2003}{005}
    [\hepph{0303210}].


\bibitem{Takahashi3}
    K. Takahashi, M. Watanabe, K. Sato and T. Totani,
    {\it  Effects of neutrino oscillation on the supernova neutrino
	spectrum},
    \prd{64}{2001}{093004}
    [\hepph{0105204}].




\bibitem{Fogli2} 
    G. Fogli, E. Lisi, A. Mirizzi and D. Montanino,
    {\it Probing supernova shock waves and neutrino flavor transitions
	in next-generation water-Cherenkov detectors},
    \newjournal{JCAP}{JCAPA}{0504}{2005}{002}
    [\hepph{0412046}].

\bibitem{Beacom:1999}
 J.~F.~Beacom and P.~Vogel,
  {\it Mass signature of supernova nu/mu and nu/tau neutrinos in
  SuperKamiokande}
  \prd{58}{1998}{053010}



\bibitem{kotake_rev} K.~Kotake, K.~Sato and K.~Takahashi,
{\it Explosion Mechanism, Neutrino Burst, and Gravitational Wave in
Core-Collapse Supernovae}
\newjournal{Rept. Prog. Phys.}{RPP}{\bf 69}{2006}{971}
 [\astroph{0509456}].

\bibitem{mikheev1985}
 S. P. Mikheev and A. Y. Smirnov,
{\it  Resonance Amplification of Oscillations in Matter and Spectroscopy
	of Solar Neutrinos},
\sjnp{42}{1985}{913} 
[\yf{42}{1985}{1441}].

\bibitem{mikheev1986}
S. P. Mikheev and A. Y. Smirnov,
{\it Neutrino oscillations in a variable density medium and bursts due
	to the gravitational collapse of stars}, 
\jetp{64}{1986}{4}
[\zetf{91}{1986}{7}].


\bibitem{balantekin}
    A. B. Balantekin, J. M. Fetter and F. N. Loreti,  
    {\it The MSW effect in a fluctuating matter density}, 
    \prd{54}{1996}{3941}
    [\astroph{9604061}] 


\bibitem{Takahashi5}
    K. Takahashi, K. Sato, A. Burrows and T. A. Thompson,
    {\it  Supernova neutrinos, neutrino oscillations, and the mass of
	the progenitor star}, 
    \prd{68}{2003}{113009}
    [\hepph{0306056}].

\bibitem{kachelriess2005}
M. Kachelriess, R. Tomas, R. Buras, H. T. Janka, A. Marek, and M. Rampp, 
{\it Exploiting the neutronization burst of a galactic supernova}
\prd{71}{2005}{063003} [\astroph{0412082}].


\bibitem{Lunardini}
    C. Lunardini and A. Y. Smirnov,
    {\it  Supernova neutrinos: Earth matter effects and neutrino mass
	spectrum}
    \npb{616}{2001}{307}
    [\hepph{0106149}]


\bibitem{dighe2004} 
 A. S. Dighe, M. Kachelriess, G. G. Raffelt and R. Tomas, 
{\it Signatures
of supernova neutrino oscillations in the earth mantle and core},
 \newjournal{JCAP}{JCAPA}{0401}{2004}{004}
[\hepph{0311172}].







\bibitem{Schirato02}
    R. C. Schirato and G. M. Fuller
    {\it Connection between supernova shocks, flavor transformation, and
	the neutrino signal }
    [\astroph{0205390}]

\bibitem{Lunardini08}
    C. Lunardini, B. M$\ddot{\mathrm{u}}$ller and H. Th. Janka,
    {\it Neutrino oscillation signatures of oxygen-neon-magnesium
	supernovae},
   \prd{78}{2008}{023016}
   [\arXivid{0712.3000}]

\bibitem{duan_rev} H.~Duan and J.~P.~Kneller,
 {\it Neutrino flavor transformation in supernovae}
  [\arXivid{0904.0974}].

\bibitem{Tomas}
    R. Tom$\grave{\mathrm{a}}$s, M. Kachelrie$\ss $, G. Raffelt, 
    A. Dighe, H. T.  Janka  and L. Scheck, 
    {\it Neutrino signatures of supernova shock and reverse shock
	propagation},
    \newjournal{JCAP}{JCAPA}{0409}{2004}{015}
    [\astroph{0407132}].


\bibitem{Takahashi1}
    K. Takahashi, K. Sato, H. E. Dalhed and J. R. Wilson,
    {\it Shock propagation and neutrino oscillation in supernova}, 
    \app{20}{2003}{189}
    [\astroph{0212195}].


\bibitem{raffelt}  
   G.~G.~Raffelt and A.~Y.~Smirnov,
   {\it Self-induced spectral splits in supernova neutrino fluxes},
   \prd{76}{2007}{081301} [\arXivid{0705.1830}].

\bibitem{duan}  
   H.~Duan, G.~M.~Fuller and J.~Carlson,
   {\it Simulating nonlinear neutrino flavor evolution}
   [\arXivid{0803.3650}].

\bibitem{dasgup1} 
B.~Dasgupta and A.~Dighe,
 {\it Collective three-flavor oscillations of supernova neutrinos}
 \prd{77}{2008}{113002}
  [arXiv:0712.3798 [hep-ph]].


\bibitem{dasgup2}
 B.~Dasgupta, A.~Dighe and A.~Mirizzi,
  {\it Identifying neutrino mass hierarchy at extremely small theta(13) through
  Earth matter effects in a supernova signal}
\prl{101}{2008}{171801}
  [arXiv:0802.1481 [hep-ph]].

\bibitem{duan2006} 
 H. Duan, G. M. Fuller and Y. Z. Qian, 
{\it  Collective neutrino flavor transformation in supernovae}, 
\prd{74}{2006}{123004} [\astroph{0511275}].

\bibitem{lim1988} 
 C. S. Lim and W. J. Marciano, 
{\it  Resonant Spin - Flavor Precession of Solar and Supernova
	Neutrinos}, 
Phys. Rev. D 37, 1368 (1988).
\prd{37}{1988}{1368}.

\bibitem{akhmedov1988}
 E. K. Akhmedov, 
{\it  Resonant Amplification of Neutrino Spin Rotation in Matter and the
	Solar Neutrino Problem},
Phys. Lett. B 213, 64 (1988).
\plb{213}{1988}{64}.

\bibitem{Akhmedov03}
    E. K. Akhmedov and T. Fukuyama,
    {\it  Supernova prompt neutronization neutrinos and neutrino
	magnetic moments. }
    \newjournal{JCAP}{JCAPA}{0312}{2003}{007}
    [\hepph{0310119}]



\bibitem{Ando03b}
    S. Ando and K. Sato, 
    {\it   Resonant spin flavor conversion of supernova neutrinos:
	Dependence on presupernova models and future prospects. }
    \prd{68}{2003}{023003}    
    [\hepph{0305052}]

\bibitem{Totani1996} 
    T. Totani \& K. Sato, 
    {\it Resonant spin-flavor conversion of supernova neutrinos and
	deformation of the electron antineutrino spectrum} ,
    \prd{54}{1996}{5975}
    [\astroph{9609035}]


\bibitem{Wang01}
    L. Wang, D. A. Howell, P.  Hoeflich and J. C. Wheeler,
    {\it Bi-polar Supernova Explosions}
    \apj{550}{2001}{1030} 
    [\astroph{9912033}].







\bibitem{marek}
 A.~Marek and H.~T.~Janka,
  {\it Delayed neutrino-driven supernova explosions aided by the standing
  accretion-shock instability}
  \apj{694}{2009}{664}
  [\arXivid{0708.33372}].

\bibitem{Iwakami07}
  W.~Iwakami, K.~Kotake, N.~Ohnishi, S.~Yamada and K.~Sawada,
  {\it Three-Dimensional Simulations of Standing Accretion Shock
	Instability in 
   Core-Collapse Supernovae},
  \apj{678}{2008}{1207}
  [\arXivid{0710.2191}].

\bibitem{kotake_mhd}
K.~Kotake, K.~Sato, H.~Sawai and S.~Yamada,
  {\it Magnetorotational effects on anisotropic neutrino emission and convection
  in core-collapse supernovae}
  \apj{608}{2004}{391}

\bibitem{Burrows07}
  A.~Burrows, L.~Dessart, E.~Livne, C.~D.~Ott and J.~Murphy,
  {\it Simulations of Magnetically-Driven Supernova and Hypernova Explosions in
  the Context of Rapid Rotation},
  [\astroph{0702539}].

\bibitem{Takiwaki09}
T.~Takiwaki, K.~Kotake and K.~Sato,
  {\it Special Relativistic Simulations of Magnetically-dominated Jets in
  Collapsing Massive Stars}
  \apj{691}{2009}{1360}
  [\arXivid{0712.1949}].

\bibitem{Burrows06}
  A.~Burrows, E.~Livne, L.~Dessart, C.~Ott and J.~Murphy,
  {\it A New Mechanism for Core-Collapse Supernova Explosions},
  \apj{640}{2006}{878}
  [\astroph{0510687}].

\bibitem{friedland}
 A.~Friedland and A.~Gruzinov,
  {\it Neutrino signatures of supernova turbulence}
  [\astroph{0607244}].
  arXiv:astro-ph/0607244.

\bibitem{choubey}
    S. Choubey, N. P. Harries and  G. G. Ross, 
    {\it  Turbulent supernova shock waves and the sterile neutrino
	signature in megaton water detectors},
    \prd{76}{2007}{073013}
    [\hepph{0703092}]


\bibitem{kneller}
J.~P.~Kneller, G.~C.~McLaughlin and J.~Brockman,
  {\it Oscillation Effects and Time Variation of the Supernova Neutrino
	Signal}, 
  \prd{77}{2008}{045023}
  [\arXivid{0705.3835}].

\bibitem{Bucciantini}
  N.~Bucciantini, E.~Quataert, J.~Arons, B.~D.~Metzger and T.~A.~Thompson,
  {\it Relativistic Jets and Long-Duration Gamma-ray Bursts from the
	Birth of 
  Magnetars},
  \newjournal{MNRAS}{MNRAS}{383}{2008}{383}
  [\arXivid{0707.2100}].

\bibitem{MacFadyen99}
  A.~MacFadyen and S.~E.~Woosley,
  {\it Collapsars - Gamma-Ray Bursts and Explosions in "Failed
	Supernovae"}, 
  \apj{524}{1999}{262} 
  [\astroph{9810274}].


\bibitem{Nagataki07}
  S.~Nagataki, R.~Takahashi, A.~Mizuta and T.~Takiwaki,
  {\it Numerical study of gamma-ray burst jet formation in collapsars},
  \apj{659}{2007}{512}.


\bibitem{harikae}
S.~Harikae, T.~Takiwaki and K.~Kotake,
  {\it Long-Term Evolution of Slowly Rotating Collapsar in Special Relativistic
  Magnetohydrodynamics}
  [\arXivid{0905.2006}].

\bibitem{piran}
T.~Piran,
  {\it The physics of gamma-ray bursts}
  Rev.\ Mod.\ Phys.\  {\bf 76}, 1143 (2004)
  [\astroph{0405503}].

\bibitem{Heger}
    A. Heger, N. Langer and S. E. Woosley, 
    {\it  Presupernova evolution of rotating massive stars. 1. Numerical
	method and evolution of the internal stellar structure},
    \apj{528}{2000}{368}
    [\astroph{9904132}]

\bibitem{Galama98}
      T. J. Galama et al.
      {\it  Discovery of the peculiar supernova 1998bw in the error box
	of GRB 980425}
      \nature{395}{1998}{670}
      [\astroph{9806175}]







\bibitem{Woosley06} 
  S.~Woosley and A.~Heger,
  {\it The Progenitor Stars of Gamma-Ray Bursts},
  \apj{637}{2006}{914}
  [\astroph{0508175}].

\bibitem{Obergaulinger}
  M.~Obergaulinger, M.~A.~Aloy and E.~Muller,
  {\it Axisymmetric simulations of magneto--rotational core collapse:
	dynamics and gravitational wave signal},
  \asas{450}{2006}{1107}
  [\astroph{0510184}]


\bibitem{neutrino}
     M. Fukugita and T. Yanagida, 2003, 
     {\it Physics of Neutrinos and Applications to Astrophysics},
     Springer, Berlin.


\bibitem{Parameter}
    M. Maltoni, T. Schwetz, M. A. Tortola and
    J. W. F. Valle,
    {\it Status of global fits to neutrino oscillations},
    \newjournal{New \ J.\ Phys.\ }{NJOPF}{6}{2004}{122}
    [\hepph{0405172}].


\bibitem{yoshida2}
    T. Yoshida, T. Kajino, H. Yokomakura, K. Kimura, A. Takamura
    and D. H. Hartmann, 
    {\it Neutrino Oscillation Effects on Supernova Light Element
	Synthesis},
    \apj{649}{2006}{319}
    [\astroph{0606042}]. 


\bibitem{threshold}
     M. B. Smy,  
     {\it Low Energy Challenges in Super-Kamiokande-III},
     \npps{168}{2007}{118}

\bibitem{hirata1988}
Hirata, K.~S., et al.\ 
{\it Observation in the Kamiokande-II detector of the neutrino burst
	from supernova SN1987A},
 \prd{38}{1988}{448}

\bibitem{nakahata}
M. Nakahata,\
{\it Supernova Neutrinos and Recent Results from Supernova-Kamiokande,}
Proceedings of {\em the Yamada Conference LIX} on 
  Inflating Horizons of Particle Astrophysics and Cosmology (H.Suzuki, J.Yokoyama,
 Y.Suto, and K. Sato, Japan, 2006, pp 29 - 37) 
 
\bibitem{burrows1992}
A. Burrows, D. Klein and R. Gandhi
{\it The Future of supernova neutrino detection},
 \prd{45}{1992}{3361}

\bibitem{beacom1999}
J. F. Beacom and P. Vogel,
{\it Can a supernova be located by its neutrinos?},
\prd{60}{1999}{033007},
[\astroph{9811350}].

\bibitem{tomas2003}
R. Tomas, D. Semikoz, G. G. Raffelt, M. Kachelriess and A. S. Dighe,
{\it Supernova pointing with low- and high-energy neutrino detectors}, 
\prd{68}{2003}{093013},
[\hepph{0307050}].


\bibitem{HK}
K.~Nakamura,
  {\it Hyper-Kamiokande: A next generation water Cherenkov detector},
  \ijmpa{18}{2003}{4053}.


\bibitem{titand}
 Y.~Suzuki {\it et al.}  [TITAND Working Group],
  {\it Multi-Megaton water Cherenkov detector for a proton decay search:
	TITAND} 
  [\hepex{0110005}].








\bibitem{raffelt2001}
 G. G. Raffelt, 
{\it  Muon-neutrino and tau-neutrino spectra formation in supernovae}, 
\apj{561}{2001}{890} [\astroph{0105250}].


\bibitem{keil2003}
 M. T. Keil, G. G. Raffelt and H. T. Janka, 
{\it  Monte Carlo study of supernova neutrino spectra formation},
\apj{590}{2003}{971} [\astroph{0208035}].

\bibitem{fogli1}		
    G.~L.~Fogli, E.~Lisi, A.~Marrone and A.~Mirizzi,
    {\it Collective neutrino flavor transitions in supernovae and the
	role of trajectory averaging},
    \newjournal{JCAP}{JCAPA}{0712}{2007}{010}
    [\arXivid{0707.1998}].

\bibitem{tanaka}
M.~Tanaka {\it et al.},
  {\it Spectropolarimetry of the Unique Type Ib Supernova 2005bf: Larger Asymmetry
  Revealed by Later-Phase Data}
 [\astroph{0906.1062}].   

\bibitem{fryer2007} 
 C. L  Fryer et al.
{\it Constraints on Type Ib/c Supernovae and Gamma-Ray Burst Progenitors},
	\newjournal{PASP}{PASP}{119}{2007}{1211}.


\bibitem{kotagw}
K.~Kotake, W.~Iwakami, N.~Ohnishi and S.~Yamada,
{\it Stochastic Nature of Gravitational Waves from Supernova Explosions with
  Standing Accretion Shock Instability}  
\apj{697}{2009}{L133}
[\astroph{0904.4300}]



\bibitem{ott2009}
C. D. Ott,
{\it The gravitational-wave signature of core-collapse supernovae},
\cqg{26}{2009}{063001} [\arXivid{0809.0695}].








\end{thebibliography}
\end{document}